# End-Users' Knowledge and Perception about Security of Mobile Health Apps: A Case Study with Two Saudi Arabian mHealth Providers

Bakheet Aljedaani[a,b], Aakash Ahmad[c], Mansooreh Zahedi[a], M. Ali Babar[a,d]

[a]CREST – the Centre for Research on Engineering Software Technologies, University of Adelaide, Australia
[b]Computer Science Department, Aljumum University College, Umm Alqura University, Makkah, Saudi Arabia
[c]College of Computer Science and Engineering, University of Ha'il, Ha'il, Saudi Arabia
[d]Cyber Security Cooperative Research Centre (CSCRC), Australia
[a,d][bakheet.aljedaani | mansooreh.zahedi | ali.babar]@adelaide.edu.au, [c]a.abbasi@uoh.edu.sa

Abstract

Mobile health apps (mHealth apps) are being increasingly adopted in the healthcare sector, enabling stakeholders such as medics and patients, to utilize health services in a pervasive manner. Despite having several benefits, mHealth apps entail significant security and privacy challenges that can lead to data breaches with serious social, legal, and financial consequences. This research presents an empirical investigation into security awareness of end-users of mHealth apps that are available on major mobile platforms. We conducted end-users' survey-driven case study research in collaboration with two mHealth providers in Saudi Arabia to survey 101 end-users, investigating their security awareness about (i) existing and desired security features, (ii) security-related issues, and (iii) methods to improve security knowledge. The results indicate that while security awareness among the different demographic groups was statistically significant based on their IT knowledge level and education level ,security awareness based on gender, age, and frequency of mHealth app usage was not statistically significant. We also found that the majority of the end-users are unaware of the existing security features provided (e.g., restricted app permissions); however, they desire usable security (e.g., biometric authentication) and are concerned about the privacy of their health information (e.g., data anonymization). End-users suggested that protocols such as two-factor authentication positively impact security but compromise usability. Security-awareness via peer guidance, or training from app providers can increase end-users' trust in mHealth apps. This research investigates human-centric knowledge based on a case study and provides a set of guidelines to develop secure and usable mHealth apps.

**Keywords**: Mobile Security, Software Engineering, Empirical Study, Mobile Healthcare.

## 1. Introduction

Mobile computing is being leveraged to offer a multitude of context-aware services, ranging from social networking to fitness monitoring and smart healthcare [1]. Mobile devices unify (i) embedded sensors (for context-sensing), (ii) installed software (to process contextual data), and (iii) wireless networking (that transmits device data) to provide context-aware pervasive services. A recent report, 'The Mobile Economy 2020', published by GSMA [2] highlighted that by the end of 2019, 5.2 billion people (approx. 67% of the global population) subscribed to mobile services with an expected increase to 70% by 2025. Moreover, in 2019 mobile systems and services generated USD 4.1 trillion of economic value (approx. 4.7% of global GDP) with further growth expected to reach 4.9% by 2024. Mobile computing is revolutionizing the healthcare sector and becoming an integral part of smart healthcare initiatives offered to end-users via mobile health applications (known as mHealth apps) [3, 4]. mHealth apps forge connections between mobile technologies and the healthcare industry to provide a variety of low cost, efficient, and digitized healthcare services such as health and fitness monitoring [5], dermatologic care [6], chronic management [7, 8], and clinical practices [9]. Research2guidance (R2G), a leading consultancy firm for mHealth technologies, reported that 78,000 new mHealth apps were added to app stores in 2017 [10] with market revenue for digital health expected to reach USD 31 billion by 2020. An ever-increasing adoption of mHealth apps by healthcare stakeholders is evident in terms of 350,000 such apps available via application repositories of Android and iOS platforms [11].

The World Health Organization (WHO) refers to mHealth as new horizons for health through mobile technologies [12] that facilitate stakeholders (e.g., governments, health units, medics, and patients) to provide or utilize health services in a pervasive, efficient, and automated manner. mHealth stakeholders rely on (a) mobile sensors to capture health-critical data, (b) mobile apps to process data, and (c) wireless networking to transmit data, regardless of geographical location or physical presence. For example, a patient with the help of a pre-installed mHealth app, without seeking prior clinical appointment, could share his/her vital signs or lab reports for consultations with medical experts across the globe. From an operational perspective, mHealth tools, technologies, and apps support the outreach of healthcare professionals, neutralizing distance, time zones, and cost factors, to provide accessible and affordable clinical practices. Despite the strategic benefits [13] and generated revenues [14], security of health-critical data is among the topmost challenges for sustainability and mass-scale adoption of mHealth apps [1, 8, 15-18]. The risks of unauthorized access to health-critical data (e.g., disease symptoms, blood pressure, and clinical reports) are on the rise due to the value of data on the 'black market' along with the socio-legal consequences of the compromised data [19]. According to a recent report by the Ponemon Institute[1], the average price to maintain each medical record increased from USD 369 in 2016 to USD 380 in 2017 due to policies, regulations, and their implementations for securing health-critical data [20]. For mobile health systems, technical features alone may not be enough to ensure security, unless they are complemented with human-centric knowledge and practices to protect critical information. For example, a study by Mylonas et al. in 2013 [21] highlights that mobile devices implement a multitude of security features including but not limited to device locks, remote data wipes, and end-to-end encryption. However, even the most sophisticated security features can never guarantee human behavior (e.g., privacy leakage, unwanted access granted) that enhances or compromises device protection and/or data security [22, 23].

Security and privacy of mHealth apps can be viewed from two different perspectives referred to as the developers' perspective (i.e., security-aware development) and end-users' perspective (i.e., security-aware usability). The developers' perspective focuses on practicing secure software development lifecycle (SDLC) to engineer apps by prioritizing security-specific requirements, implementing data encryption methods, and performing vulnerability testing. For example, a recent empirical study by Aljedaani et al. in 2020 [24] focused on the developers' perspectives about critical challenges, recommended practices, and motivating factors to develop secure mHealth apps. Developers who practice secure SDLCs to engineer mHealth apps often assume that the delivered app is secure; however; end-users of the app in many instances may find security features hard to understand, get deceived by hackers and leak private information, or be mislead by app permissions to disclose classified data. One study by Atienza et al. in 2015 [23] engaged 24 focus groups with more than 250 participants to investigate the attitudes and perceptions of app users regarding mHealth systems being used at healthcare centers. Existing research indicates that sophisticated cyber-attacks jeopardize mobile apps and implemented security features often become obsolete due to state-of-the-art techniques for data infiltration/exfiltration [1, 8, 15]. To ensure security and ultimately strengthen the confidence of stakeholders in adopting mHealth systems, there is a need to unify both the developers' and end-user's perspectives of security. A recent mapping study on secure and private mobile health systems, conducted by Iwaya et al. in 2020 [25] highlighted that there is little research on understanding and measuring security awareness of end-users of mobile and ubiquitous health systems.

To empirically investigate the security-awareness of end-users[2] towards using clinical mHealth apps, we conducted end-users' survey-driven case study research by collaborating with two of the largest mHealth providers in Saudi Arabia. In our context, security awareness of end-users refers to *human-centric perception of (existing and desired) security features, experiences with security issues, and understanding of methods for secure usage of mobile health systems.* For example, biometric verification of users by an app is perceived as a security feature, excessive or undesired permissions (e.g., reading contacts or voice data) is considered as a security issue, while security training (e.g., electronic material or workshops) enables end-users to improve security-awareness. End-users included patients and medical professionals with diverse experience of clinical practices in varying roles including but not limited to medical doctors, nurses, and healthcare supervisors. Demographic analysis of end-users' data highlighted their educational backgrounds, IT skills, and years of experience using mHealth systems compatible with major mobile platforms, including Android and iOS. To objectively measure and understand security awareness, we identified three Research Questions (RQs) to be investigated:

*RQ-1: To what extent are the mHealth app users aware of the existing security features? and what are the security features that mHealth app users desire to have in mHealth apps in the future?*

**Security Perception:** The objective of this RQ is to investigate the security-awareness in terms of understanding of the existing features and desire for futuristic features that enable or enhance app security.

---

[1] Ponemon Institute available at https://www.ponemon.org/

[2]We use the term *end-users*, *participants*, *respondents* interchangeably throughout the paper, all referring to *users* of mHealth apps that were engaged in this study for their feedback and responses.

*RQ-2: What security issues have been faced by users during their usage of the security features within mHealth apps?*

**Security Challenges:** The objective of this RQ is to identify and analyze various security-related challenges faced by end-users and their impact on data privacy and apps' usability.

*RQ-3: What methods help end-users to improve their security knowledge regarding the usage of mHealth apps?*

**Security Knowledge**: The objective of this RQ is to understand methods and techniques that facilitate end-users to increase their knowledge about secure usage of the apps.

In order to answer these RQs, we gathered data from 101 respondents through a survey questionnaire and synthesized the survey data using statistical methods in two different phases. Data analysis included (i) descriptive analysis to investigate end-users' responses and (ii) qualitative analysis to identify, analyze, and report frequent overlaps in users' responses, representing patterns of security awareness. The results highlight significant variation in end-users' security-awareness. Factors such as educational background, level of IT knowledge, and prior experience with mHealth apps reflected positive impacts on security awareness. End-users perceive controlled app permissions, user authentication and security customization as useful existing features and they desire usable security along with preservation of privacy as missing but required features of their mHealth apps. Security protocols such as excessive permission requests or multi-stage authentication enhance app security but hinder app usability. A lack of security education and training by mHealth app providers contributes to a reluctance to adopt mHealth apps. We outline the primary contributions of this research as:

- Reporting an exploratory case study that investigates users' perceptions, issues, and knowledge about security of mHealth apps. The results of our study provide a taxonomy and a benchmark to evaluate the effectiveness of security features offered by mHealth apps.
- An empirically derived set of guidelines to facilitate researchers, practitioners, and stakeholders to develop and adopt secure and usable mHealth apps for clinical practices and public health.

The rest of this paper is organized as follows. Section 2 describes the background of our study. Section 3 presents related work. Section 4 details the research methodology. Sections 5 to 7 report study findings. Section 8 discusses key findings of the study. Section 9 describes validity threats for this study. Section 10 concludes the paper.

## 2. Background

We discuss the background of this study and highlight various security-related concerns for mHealth apps based on the illustrations in Figure 1. We provide an overview of the security of mHealth systems (Section 2.1) and discuss developers' and end-users' perspectives on mHealth apps (Section 2.2). The concepts and terminologies introduced in this section are used throughout the paper to guide the study design and results.

### 2.1. Security and Privacy of mHealth Systems

The app repositories provided by the world's leading and most adopted mobile platforms (i.e., Android and iOS with a joint market share of 99%) offer a multitude of mHealth apps and systems [10]. These apps vary, based on the type of mHealth services they offer and the granularity of health-critical data they collect, process, and exchange [11]. mHealth apps generally include, but are not limited to, health and fitness monitoring, medical consultations, clinical services, and medical imaging. As in Figure 1, despite the classification and healthcare services they offer, all of the mHealth apps rely on capturing personal data (age, gender, location etc.) and health-critical information (such as body temperatures, vital signs, and disease symptoms) that can be vulnerable to various security threats. Considering the context of mobile computing in general and mHealth systems in particular, security threats arise when attackers or malicious agents exploit existing vulnerabilities found in the operating system or in any third-party applications to gain access to device resources and data [4]. Specifically, any tempering, selling to third parties, or breaching individual's privacy of health-critical data can have serious social, legal, and financial consequences for both the app users and providers. Recently, some policies and regulations such as the European General Data Protection Regulation (EU GDPR) [26, 27] and the Health Insurance Portability and Accountability Act (HIPAA) [28, 29] enforce constraints on systems and ensure transparency of mechanisms that collect private data of end-users. Therefore, security of mHealth apps becomes a significant issue due to the privacy and integrity requirements of health-critical data and the regulations to ensure that privacy and integrity of personal data is maintained [5, 30].

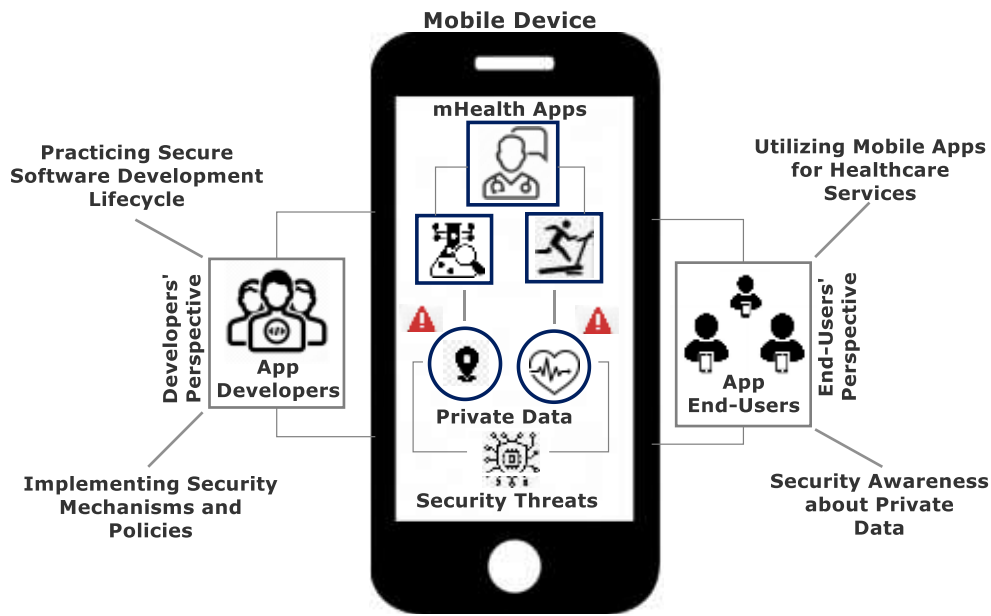

Figure 1. Overview of the Developers' and End-Users' Perspectives on Secure mHealth Apps

*Security vs. Privacy:* Data security and privacy are often considered as virtually synonymous terms, used interchangeably, referring to securing and/or protecting critical or classified information [25]. In mHealth systems, security and privacy of health-critical data are complementary concepts [25, 31]; however, for technical reasons, a distinction between the two must be maintained. Specifically, security refers to the implementation of practices and processes that ensure the confidentiality, availability, and integrity of health-critical data by restricting data usage or access by unauthorized entities [32]. In comparison, privacy has no absolute definition as it represents the basic human right (i.e., end-users' determination) about what, how much, when, and with whom to share information that is deemed as private to an individual (or a group) [33]. For example, in an mHealth app, privacy allows a user the right to grant or deny any access to their location, voice, contacts or health data whereas security mechanisms such as anonymization, encryption, or data blockage enables privacy preservation. In this research, we focus on security of mHealth apps where privacy becomes implicit such as implementing security mechanisms to preserve/protect the privacy of the users.

## 2.2. Developers' and End-Users' Perspective to mHealth Applications

Developers and end-users represent two actors with distinct but complementary roles to support the secure development and usage of mHealth apps, as illustrated in Figure 1. Developers focus on the application of secure SDLC to engineer their apps that are secure and usable for end-users. Moreover, developers working in an individual capacity or as part of development teams are responsible for implementing the required security mechanisms (e.g., encryption, authentication, secure storage, and access control) and privacy policies to enhance the confidentiality and integrity of health-critical data [26-29]. A recently study [24] engaged a total of 97 mHealth app developers to investigate the challenges, motivating factors, and best practices to develop secure mHealth apps. The study provides a set of guidelines, and SDLC practices to develop security-aware apps, but highlights that even the most advanced security features may not be sufficient if end-users' lack security awareness about protecting their private information. As shown in Figure 1, end-users utilize the apps and developers most often assume that users' have appropriate security-awareness (e.g., how to enable biometric authentication) regarding the use of mHealth apps. Security awareness enables end-users to understand the potential security threats (e.g., privacy leakage) and enable available countermeasures (i.e., reject excessive app permissions) to enhance the security and privacy of health-critical data. Lack of security awareness might lead to granting more permissions than necessary to unintentionally share health-critical data or allow other apps to unnecessarily access it [4]. Contrary to the developers' view [24], we focus on the end-user perspective in Figure 1, and it is important to measure their security awareness when using mHealth apps. The findings help to benchmark and evaluate the effectiveness of mHealth apps security features that need to be considered by developers while engineering the existing or next generation of mHealth apps.

## 3. Related Work

The related work on this topic can be generally classified into two categories, i.e., (i) end-users' security awareness (in Section 3.1) and (ii) end-users' awareness towards privacy policies for mHealth apps (in Section 3.2). Table 1 helps to objectively compare our research in terms of its scope and contributions in the context of existing work.

*3.1. End-users' Security Awareness towards mHealth Apps*

In contrast to general purpose mobile applications, mHealth apps collect, process, and exchange a multitude of private data that can be vulnerable to various security threats such as tempering of medical records, data sniffing for targeted advertisements, and identity theft [1]. For example, the magnitude and variety of private data including personal details (e.g., age, gender, location coordinates etc.) and health-critical information (e.g., disease symptoms, clinical reports, medical prescriptions etc.) make it challenging for end-users to protect sensitive information from undesired access [16, 34]. From the perspective of mHealth app providers, in addition to ensuring that secure mHealth apps have been delivered by developers, end-user training and security-awareness must be treated as a priority before deploying and operationalizing mobile health systems [4]. On the contrary, some recent studies highlight that a lack of knowledge or understanding of end-users regarding security features is still being overlooked as a threat [4, 18, 35, 36]. In the context of mHealth SDCL, prime importance is given to the development of security-aware apps with developers and mHealth providers having a collective assumption that the delivered app is secure. Despite the fact that app(s) delivered by following a secure SDLC can have well-implemented security features, for end-users such complex implementations could be either hard to understand or utilize, or require human intervention to operate [37]. As a typical example, to avail-of nearby healthcare services or consulting available medics, end-users may end up providing excessive permissions such as current location, activity, or active social contacts without being notified [4]. Understanding the security preferences of end-users can help app developers and providers to maintain the required balance between security and usability of mHealth apps [22].

*Security perceptions of end-users:* End-users' security perception of using mHealth apps can vary based on the type and context of data that is handled by the apps. Specifically, a study by Atienza et al. [23] reported that end-users' perceptions and attitudes towards the security of mHealth apps are highly contextualized based on the type of data collected by an app, time and context of access, i.e., who accessed the data, at what time they accessed it, and why. Alternatively, some end-users do not mind sharing their health-specific data on social networks, gathered by health and fitness monitoring apps (e.g., workout, walking distance, and burnt calories). An empirical study used mixed methods to collected qualitative data using six focus groups of 44 end-users and interviews of five individuals [38]. Most of the study's participants affirmed that one of the barriers to continuing to use mHealth apps is sharing personal information that might be exploited by a third party (e.g., insurance companies or advertisers). Such data sharing, despite having social implications, is perceived as volunteering efforts to encourage others to engage in a healthy routine and lifestyle or obtain support and feedback from social contacts. In another study [22], through structured interviews, the authors sought to identify the desired features of mHealth apps which would enhance the trust of end-users. The participants agreed that their confidence level increases when using mHealth apps that enable end-users to adjust security settings, enforce regular updates for passwords, and allow monitoring data access via access logs. Moreover, the participants suggested biometric verification, providing end-to-end-encryption for data-in-transit and data-at-rest, and allowing end-users to wipe the device remotely, once it is lost or stolen. The study in [39] investigated whether a brief security education offered in a mHealth app (i.e., Security Simulator app) can change end-users' security behavior in terms of choosing appropriate security settings. The participants were asked to make security selections in the developed app before and after they viewed the consequences of security features. The participants' selections before and after the security education were compared to determine the effectiveness of security education to improve security awareness. The findings of the study concluded that simulation-based education is helpful in changing end-users' security behavior and helps them to select stronger security measures.

*Use of personal devices for mHealth systems:* In clinical setting environments, health practitioners are often encouraged to use their own devices (i.e., Bring Your Own Device - BYOD) [36]. Personal devices can be customized and convenient for practitioners to work with mHealth systems; however, such devices lack strict authentication or lock mechanisms, which make end-users' data vulnerable to undesired access. Considering the BYOD scenario, healthcare professionals are usually granted more access privileges as administrative users of mHealth apps. Therefore, when such professionals run other mobile apps that frequently access device data, in parallel to mHealth apps, it can compromise the security and privacy of health-critical data. Modern devices with up to date security patches are equipped with security features including device lock mechanisms, end-to-end encryption and remote data wiping for enhanced protection [21]. In a recent survey study, more than 450 smartphone owners clearly indicated that they did not use the security features offered by the devices such as frequent password updates, biometric verification or mechanisms for controlled data access [22].

*3.2. End-users' Awareness of Privacy Policies for mHealth Apps*

Privacy policies represent a set of legal statements put forward by regulatory bodies. These regulatory statements must be incorporated in mHealth systems to make data collection, processing, and transmission transparent to a data contributor, e.g., an end-user [40]. As a standard practice, privacy policies are presented to end-users before the installation of mHealth apps. A lack of awareness of end-users about privacy policies can be mainly due to (i) mHealth apps providers lacking clarity and transparency in presenting such policies or (ii) end-users themselves overlooking or failing to read through such policies to understand their consequences [35, 36]. A study by Parker et al. in 2019 [36] conducted content analysis to investigate privacy issues for 61 mental health apps. The results of the study suggest that in most cases, privacy policies are ambiguous and there is a need to provide information to end-users about how and when their health data is collected, retained, or shared with any third parties. The study found that most of the analyzed apps encourage end-users to share their health-critical data on social networking platforms without rationalizing the social and legal consequences [36]. Moreover, in many cases, privacy policies are hard to read and understand due to the use of complex language and notations. Another study [41] performed content analysis of consumer perspectives on mHealth apps for bipolar disorder patients in particular. The study's participants showed that they have no problem with upgrading apps or buying some features that can help to maintain their privacy. There is a need to increase privacy-awareness of end-users and to achieve that mHealth providers should support explicit presentation of privacy policies as part of user training [35, 36]. We assert that privacy policies should answer some fundamental questions about end-users' rights regarding their private information, such as how to terminate data collection, how to remove health data from an app's servers, and how and to whom end-users can complain [18, 22, 36, 42]. In the absence of explicit presentation and users' training regarding privacy policies, even the advanced security features cannot protect the integrity of personal information and health-critical data [39].

**Conclusive Summary:** We now provide a conclusive summary and comparative analysis of the most relevant existing research with the proposed study in Table 1. To compare, we classify the most relevant research among three categories namely (a) security-aware usage [22, 23, 43], (b) security policies for secure usage [38, 39], and (c) security-aware development of mHealth apps [24], as in Table 1. Comparative analysis is based on four criteria in Table 1: (i) *research challenge(s)*, (ii) *focus and contributions*, (iii) *evaluation context*, and (iv) *research limitations*. The *study reference* points to an individual research work under discussion and its *year of publication*. For example, a study [24] published in 2020 aims to address the challenges pertaining to developers' views on secure SDLC for mHealth apps. The study focuses on challenges, practices, and motivators for secure development of mHealth apps. The research was evaluated based on a survey of 97 mHealth app developers; however, the small number of participants and the single source of data (i.e., web-based survey of respondents) represents study limitations [24]. Based on the data in Table 1, to the best of our knowledge, there has been no empirical study to investigate the end-users' perspectives and their security awareness toward using clinical mHealth apps. The scope of our study is precise in terms of investigating clinical mHealth apps such as patient management systems that handle highly sensitive health-critical data and personal information. Our study is expected to provide an in-depth view of the security awareness of end-users through evaluating the relationship between end-users' security awareness and their characteristics and by identifying specific security features that may enhance end-users' confidence in using mHealth apps, as in RQ1. We also aim to determine the various security-related challenges faced by end-users and their impact on data privacy and the usability of apps, as in RQ2. Furthermore, we present the implemented methods to ensure end-users have the underlying knowledge of the mHealth apps they use, as in **RQ3**.

Table 1. Comparative Analysis of Most Relevant Existing Studies Compared to our Study

| Study Reference | Research Challenges | Focus and Contributions | Evaluation Context | Research Limitations | Year of Publication |
|---|---|---|---|---|---|
| Empirical Studies on Security-aware Usage of mHealth Apps ||||||
| [43] | Investigated the security awareness of end-users of mHealth apps. | - Security knowledge<br>- Security attitude<br>- Motivating behaviour | - End-users Survey (101 participants)<br>- Quantitative Analysis | - Bias in data collection (Survey only)<br>- Limited to End-users from one region. | 2020 |
| [23] | Investigated the security and privacy of mHealth apps from end-users' views. | - Users' attitude for security<br>- Users' concerns for security | - Focus Groups (256 participants) | - Limited to functionality of apps<br>- Source of data collection | 2015 |
| [22] | Investigated security barriers for end-users of mHealth apps. | - End-users' security and privacy<br>- Desired security features | - User Survey (117 participants)<br>- Focused Group Interviews | - Diversity of participants<br>- Bias in data collection | 2019 |
| Investigating Impacts of Security Policies on Usage of mHealth Apps ||||||
| [39] | Methods to encourage end-users of mHealth apps to follow stronger security measures. | - Impacts of security settings<br>- Simulating security scenarios | - User Survey (66 participants)<br>- User Interviews | - Limited to specific group of end-users, i.e., young and highly educated (Bachelor's degree or higher). | 2018 |
| [38] | Empirical analysis of end-user's perceptions towards using mHealth apps. | - Usability patterns of app<br>- Users' security knowledge | - Focus Groups (44 participants)<br>- Individual Interviews (5 participants) | - Lack of app usage experience<br>- Number of Participants | 2016 |
| Survey of Security-aware Development of mHealth Apps ||||||
| [24] | Investigated developers' perspective on secure SDLC for mHealth Apps. | - Security challenges for SDLC<br>- Security practices for SDLC<br>- Motivating factors for SDLC | - Developers' Survey (97 participants)<br>- Qualitative Analysis | - Number of participants<br>- Source of data collection (Survey only) | 2020 |
| Proposed Study | Empirically investigate security awareness of end-users of mHealth apps (clinical settings) | - Security perception (*existing vs. desired features*)<br>- Security challenges (*security vs. usability*)<br>- Security knowledge (*self vs. training*) | - End-users' Survey (101 participants)<br>- Qualitative Analysis<br>- User perception taxonomy | - Diversity of participants<br>- Source of data collection | N/A (conducted in 2020) |

## 4. Research Design

Since we targeted two specific populations to find answers for the outlined RQs, we conducted an exploratory case study to seek an understanding of end-users' security knowledge and perception towards using clinical mHealth apps [44]. A case study is defined as an empirical method that aims to investigate contemporary phenomena in their context [45]. Our case study was conducted through a survey following the research protocol that we developed in advance. In this section, we discuss the research methodology, which is comprised of three phases. Each phase is detailed below as per the illustrations in Figure 2.

*4.1. Phase 1 – Design Case Study Protocol*

The results of conducting a rigorous literature review (Section 3) helped us to analyze the collective impact of the existing research, highlighting proposed solutions and their limitations, to specify the research questions and design the survey questionnaire as in Figure 2. In the study protocol, we (i) *specified the research questions* (RQ1-RQ3 outlined in Section 1) (ii) *designed the survey questionnaire* (presented in **Appendix 3**), and (iii) *identified the data collection methods*. The findings of the literature review and the guidelines provided by Kitchenham and Pfleeger in [46] helped us to design the survey questionnaire that collected end-users' perspectives on mHealth app security. As highlighted in Figure 2, the survey (available at **Appendix 3**) contained a total of 13 Questions (**Q1-Q12** and one **optional**) to capture end-users' feedback and recommendations.

*Survey Questionnaire for End-users*: In order to provide a fine-grained analysis of end-users' feedback and to answer each of the RQs independently, we categorized the survey questionnaire into seven different categories. First, a series of questions (**Q1-Q7**) which record the demography of end-users to help correlate end-users' security knowledge based on factors such as age, level of IT knowledge, and education. **Q8** captures users' perspectives about the importance of securing health-critical data, presented as a multi-choice option via a Likert scale to (i.e., *very important*, *important*, *neutral*, *not important*, *not important at all*). **Q9-Q10** capture end-users' awareness about the existing security features and security features that are desired as an added value to secure mHealth apps. Specifically, **Q9** presents eight security features identified from the literature using a Likert-scale with five options for each feature (i.e., *always*, *sometimes*, *rarely*, *never, and I don't know*). **Q10** is an open-ended question that asks end-users to suggest the security features that may be missing from the existing apps but are highly desired. **Q11** is an open-ended question that aimed to explore the security barriers experienced by end-users during their usage of mHealth apps. **Q12** is an open-ended question that aimed to find out the methods used to make end-users aware of the security of mHealth apps. An optional question is also added at the end of our survey to allow respondents to share their comments and/or feedback.

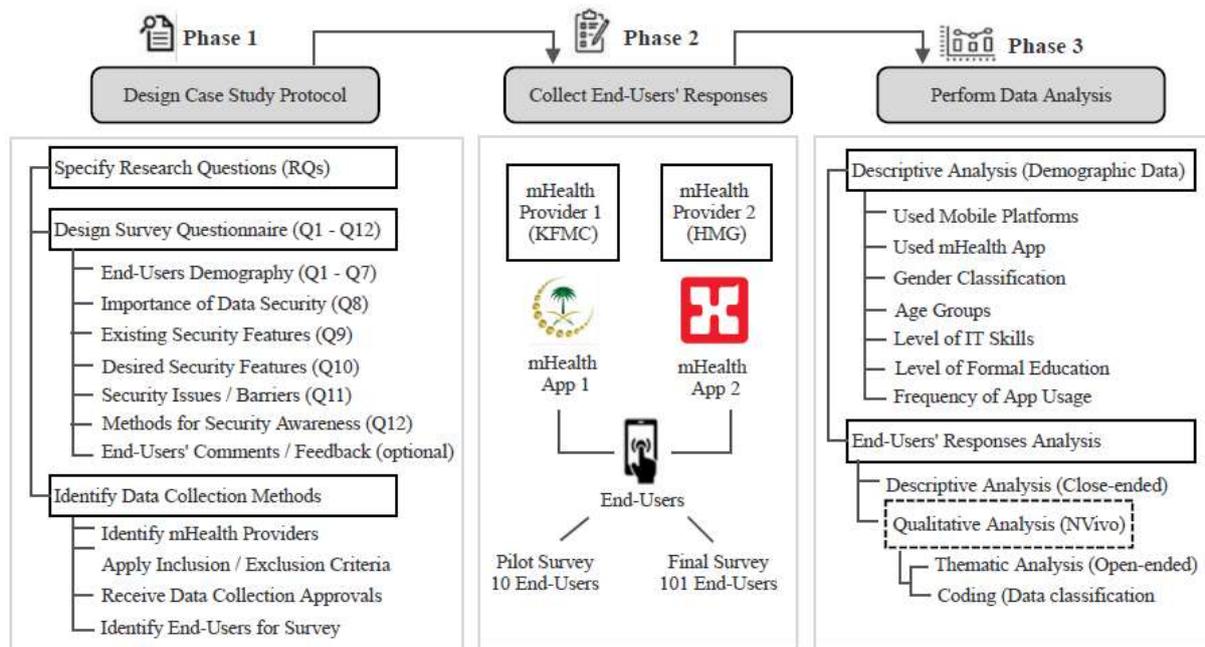

Figure 2. Overview of the Research Methodology

*Selection of mHealth apps*: To identify which mHealth apps to investigate and how to collect data for the study, we searched in the two major app repositories (i.e., Google Play by Android and App Store by iOS) to find out which mHealth apps are available and who provides them. We needed to conduct a case study with mHealth app providers that (i) have functional mHealth apps and (ii) could provide us with a survey of their end-users (medical professionals, patients, clinical technicians, etc.). We were interested in the type of apps that can be used in the clinical setting by enabling their users to access a wide range of services (e.g., reviewing medical records, viewing scan images and lab results, etc.). We limited our search for mHealth apps, provided by approved Saudi Arabian mHealth providers. Saudi Arabia can be considered as one of the fastest developing countries and rapidly adopting mHealth apps for clinical practices and public health initiatives. Based on our search of the available mHealth apps that have been provided by health providers, we found 16 mHealth apps. We excluded 4 health providers since their apps had low number of downloads or did not interact with end-users for their health data. We contacted 12 health providers to seek their approval to collect data from on-site. Eventually in December 2019, we received approval from two health providers[3], namely (i) *King Fahad Medical City* – KFMC (with the iKFMC app, launched in 2017, which has more than 50,000 downloads, 420 reviewers, and a user rating of **3.8**), and (ii) the Dr Sulaiman Al Habib Medical Group – HMG (with the Dr. Sulaiman Alhabib app, launched in 2016, which has more than 100,000 downloads, 9034 reviewers, and a user rating of **4.4**).

### 4.2. Phase 2 – Collect End-users' Responses
To collect end-users' responses, the two mHealth providers (KFMC, HMG) were purposefully chosen because they provide mHealth apps which allow their users to access a wide range of services such as creating, storing, and sharing medical records, viewing scanned images, lab results, and automating their clinical practices. The first author personally visited both mHealth providers during January and February 2020 to conduct the case study and collect data. The end-users willing to participate in the survey were provided with the option to either take a hard copy to be filled out or scan a QR code with their devices to access the online version of the survey. The survey preamble briefly described the purpose of the study, and who would be eligible to participate. To maintain the reliability of the collected responses, we focused on ensuring that potential respondents were: 1) briefed about the survey and one researcher was present at all times to clarify any issues, 2) allowed to exit the survey at any time they desired, and 3) experienced in using the provided mHealth apps. As in Figure 2, first, we conducted a pilot study with approximately 10% of the respondent population which helped us to finalize the survey questionnaire. Incomplete responses were removed, and thus, a total of 101 accurate and acceptable responses (referred to as **R1** to **R101**) from the end-users were collected on-site at KFMC and HMG. The survey took approximately 10 minutes to complete.

---
[3] King Fahad Medical City (KFMC) https://www.kfmc.med.sa/EN/Pages/Home.aspx
mHealth app: *iKFMCApp*: https://play.google.com/store/apps/details?id=sa.med.kfmc&hl=en
Dr. Sulaiman Al Habib Medical Group (HMG) https://hmg.com/en/pages/home.aspx
mHealth app: Dr. Sulaiman Alhabib App: https://apps.apple.com/ae/app/dr-sulaiman-alhabib/id733503978

It is worth mentioning that all end-users that we surveyed and both the mHealth providers were based in Saudi Arabia, thus limiting the geo-distribution of survey participants and its impact on study findings. Some travel restrictions globally, during the said time period (i.e., January and February 2020), also limited our planned on-site data collection from more countries. Extended work is in progress designing and deploying a web-based survey and a case study - to seek feedback from geographically diverse end-users and their healthcare providers.

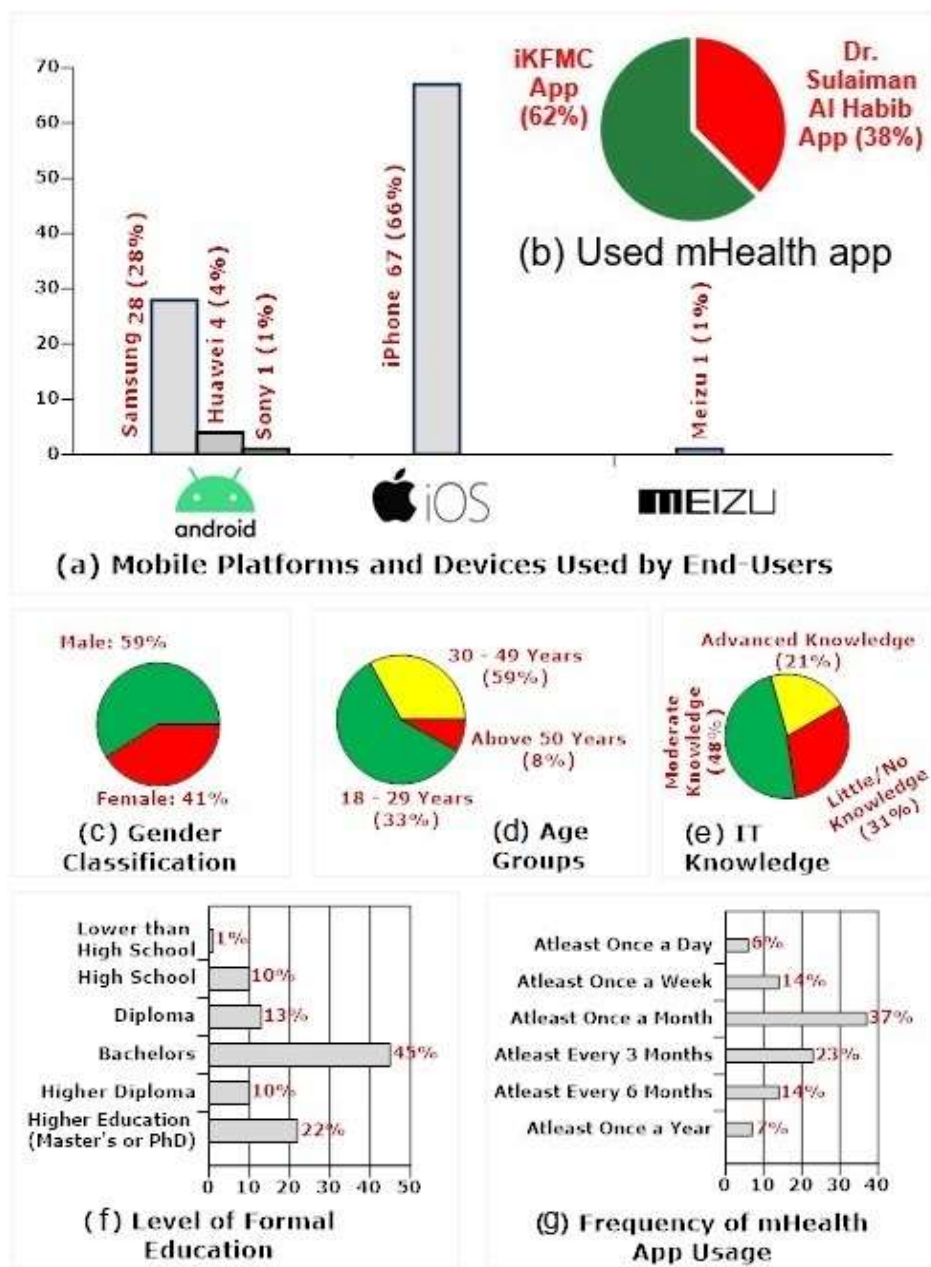

Figure 3. Demographic Details of mHealth Apps End-Users (Sample Size =101)

*4.3. Phase 3 – Perform Data Analysis*

In the last phase of the methodology, we performed data analysis based on end-users' responses as in Figure 2. We used SPSS version 27 (IBM) for the quantitative analysis of data. A descriptive analysis was conducted to report the respondents' demographic data. As in Figure 3 above, the demography analysis included factors including but not limited to, mobile platforms and mHealth apps used, gender classification, age group, and level of end-users' IT knowledge. The respondents' demographic information was used to contextualize the responses that complement security-awareness for a specific group of users. For example, we were able to investigate if the *level of formal education*, *IT knowledge* and *specific mobile platforms* had an impact on the security awareness of end-users. To determine the security awareness regarding the existing security features for a specific group of users, we conducted the Independent-Sample T-test for gender since we were comparing two independent populations (i.e., male and female), and the Kruskall-Wallis H test for more than two independent populations (e.g., IT knowledge level, age group, etc.) followed by a post hoc test to examine the significance of differences in the mean scores for the specific group of users. For each demographic data, we tested the null hypothesis (i.e., *H0: there is no significant difference*) against the alternative hypothesis (i.e., *H1: there is a significant difference*), whereas $\mu1, \mu2, ..., \mu k$ refers to population means. Since including the statistical analysis within the results made the results section quite lengthy, we included our statistical analysis in Appendix 1 and the full reports for each statistical test in [47]. Furthermore, a descriptive analysis was used to report the participants' responses to the Likert-scale questions (**Q9**) that investigate the end-users' security awareness of existing security features (i.e., the first part of **RQ1**).

For qualitative data, a thematic analysis method [48-50] was performed to analyze the textual responses of end-users (i.e., **Q10** - **Q12**) corresponding to **RQ2**, **RQ3**, and the second part of **RQ1**. To further enhance the analysis, NVivo[4] was used to organize and analyze the data as it provides a convenient mechanism for comapring the emerging themes. The initial themes were done by the first author and reviewed and revised (wherever required) by the second author to avoid potential bias. We mainly followed the five steps of the conceptualized thematic analysis method. First, we reviewed and examined the provided responses to determine the parts that were relevant to each of the research questions indicated in section 1. Second, we derived the initial codes for each research question. Third, we tried to combine different initial codes generated from the second step into potential themes. Fourth, we reviewed and refined the identified themes by checking them against each other to understand what themes had to be merged with others, or dropped. Lastly, we assessed the trustworthiness of each main theme and created a name for each of them. Figure 4 demonstrates an example of the qualitative data analysis that led to our findings. For example, as in Figure 4 in the context of RQ-1 (end-users' understanding of existing security features and the desired security features) one of the participants' responses was, *'Use of finger print is helpful to login but sometimes verification text arrives too late ... '*. This response helped us to identify that biometric based 2FA, which follows an SMS activation code, is a desired secuirty feature by end-users; however, any other issues that delay the reception of the activation code can be an issue as they could restrict the user from logging in to the app. Figure *4* also highlights that end-users' responses related to app usability such as *'... using distinct colours and sound notifications will help'*, were discarded from the analysis to strictly focus on security aspects of the apps.

*Ethics Approvals:* Both mHealth providers granted approval for on-site data collection through their Institutional Review Board (KFMC approval number: *19-462E*, HMG approval number: *HAP-01-R-082*). Our study was also approved by the Human Research Ethics Committee at the University of Adelaide (H-2019-165). Further details of the methodology, statistical data analysis, and ethics approvals are available in [47].

---

[4] https://www.qsrinternational.com/nvivo-qualitative-data-analysis-software/home/

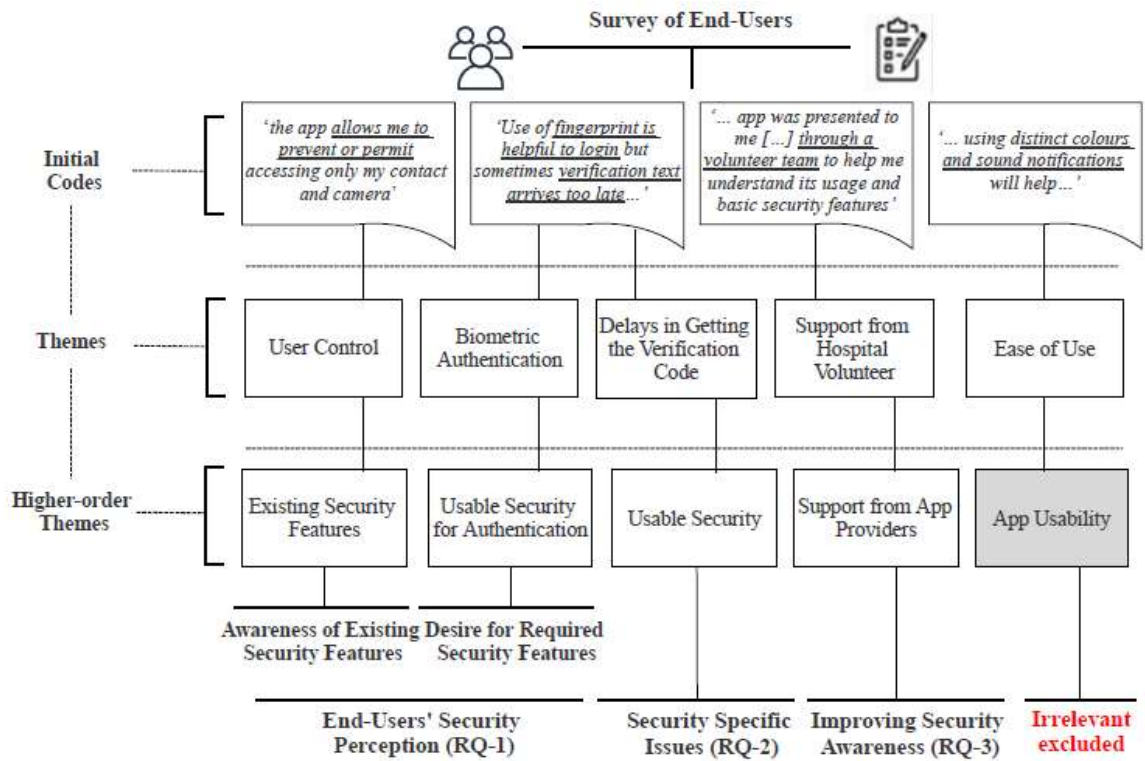

Figure 4. The Steps of Applying Thematic Analysis on Qualitative Data

**Findings of the study:** We now present the findings of the study to answer the outlined RQs. Answers to the RQs are presented in the dedicated sections focused on end-users': (i) *security perceptions* (**RQ1**) in Section 5, (ii) *security issues* (**RQ2**) in Section 6, and (iii) *security knowledge* (**RQ3**) in Section 7. Technical details of the statistical analysis and hypotheses testing about security awareness in the context of end-user demography (Figure 3) are presented in **Appendix 1** (Table 4 – Table 5). For the sake of illustration, Table 2 complements the presentation of the study results (**RQ-1**) by exemplifying end-user responses, denoted as [**Rn**], where n represents a numerical value ranging from 1 to 101 for unique identification of each response. Examples for end-users responses for **RQ-2**, and **RQ-3** are presented in **Appendix 2** (Table 6 – Table 7). For illustrative reasons, we provide visual markings, wherever required, indicated  as to express 'User Response'  and to express conclusive summary based on data analysis.

## 5. End-users' Understanding of Existing Security Features and the Desired Security Features (RQ1)

In this section, we answer RQ1 which aims to investigate security-awareness in terms of understanding of the existing features and desire for futuristic features that enable or enhance app security. Some recent studies, such as [22], have suggested that users' perception about security is based on their (a) knowledge of the existing features (i.e., available security measures) and (b) understanding of the desired functionality (i.e., required security measures). The questionnaire-based study, which contains open-ended questions, and closed questions, is a suitable method to capture the knowledge of end-users, as in [22]. Therefore, we investigated the security based on users' awareness about existing features using closed questions, as in Section 5.1, and recommendations about the desired features that can optimize security measures and strengthen users' trust in the app using open-ended questions, as in Section 5.2.

### 5.1. Security Awareness about Existing Features

To understand security awareness regarding the existing features (**Q9**, per survey design in Figure 2, Phase 1), we formulated eight Security Statements referred to as **SA1 – SA8**, visualized in Figure 5. The statements were formulated based on identifying eight distinct security features provided in the chosen mHealth apps (iKFMC app, Dr. Sulaiman Alhabib app, from Figure 2). The statements were divided into four categories namely (1) App Permissions [**SA1, SA2, SA3**], (2) User Authentication [**SA4, SA5**], (3) User Control [**SA6, SA7**], and (4) Feedback and Reporting [**SA8**]. The statements were classified into categories based on their relevance to objectively assess: *how end-users perceive security (and privacy) of their health-critical data while using the provided features by mHealth apps*. For example, the statement **SA1–SA3** that corresponds to users' consent and permission to share their data was organized under the App Permissions category. Each of the eight statements was presented as a five scale Likert option (i.e., *always*, *sometimes*, *rarely*, *never, and I don't know*) to determine users' awareness of an existing feature.

Regarding/ In terms of the apps, the existing security features refer to implemented controls, policies, and procedures (e.g., authentication protocols, users' permissions, data wiping) to enable app security. Table 2 provides examples of users' responses corresponding to **SA1** – **SA8**. An overview of the results is illustrated in Figure 5. The uncertainty in users' responses is reflected through their selection of the options *sometimes* and *rarely*. We exemplify some responses (**SA3**, **SA7**) to explain App Permissions and User Control features where end-users indicated a lack of security perception as: *'the app sometimes collects data without (my) permission* [**R43**]'. *'the app rarely provides the feature of wiping all health data in case the device is lost or stolen'* [**R4**].

The results indicate that more than 40% of the respondents suggested that they were unaware of app permission to access their private data, whereas only about 20% indicated they lacked knowledge about wiping their health-critical data. It is important to mention that an overwhelming majority, i.e., more than 83% of respondents, suggested that they were aware that apps did not collect more personal information than required (SA2). For example, as in Table 2, the response [**R36**] indicated that end-users perceived the app as easy to use for authentication and did not collect excessive information.

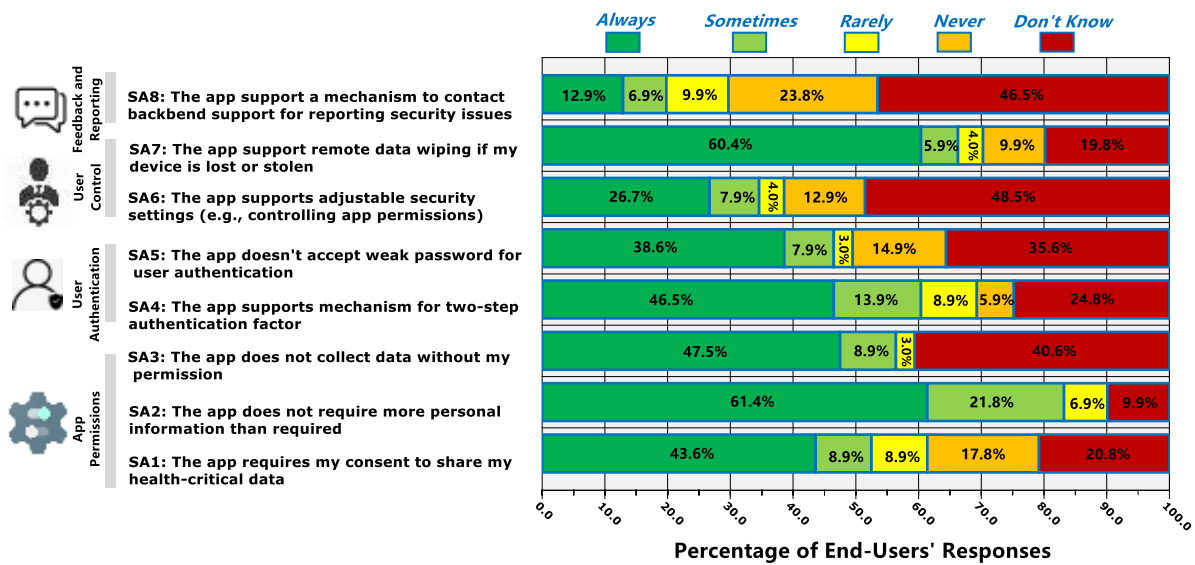

Figure 5. End-Users Awareness about Existing Security Features (n=101)

Overall, it is believed that end-users still lack awareness about the implemented security features within the apps. For example, 48.5% of our respondents do not know whether the provided apps contain adjustable security settings or not (**SA6**). Also, 46.5% do not know whether the examined apps have the feature of contacting backend support to report security issues (**SA8**). These percentages can be considered very high in representing nearly half of the surveyed respondents. Thus, further support is needed to familiarize end-users with current security features, when and how they could be utilized.

## 5.2. Desired Security Features in mHealth apps

To understand the second dimension of end-users' security awareness, i.e., the desired security features, we presented an open-ended question (**Q10,** per survey design in Figure 2, Phase 1). The question inquired about the security features desired by the users in existing mHealth apps. The open-ended question aimed to capture and compile a wish list that users perceive as vital to further optimize the security features of mHealth apps they use. Based on users' responses, we identified two desired features, namely usable security and privacy preservation that can be further classified into nine sub-features, as in Figure 6. The text from users' responses was analyzed to identify recurring themes, i.e., repeated text patterns for classification and generic naming of the features (illustrated in Figure 4). Once the generic features were identified, we put specific features of users responses under fine-grained analysis, discussed in the subsections below. Figure 6 complements the discussion of the desired security features based on a relative percentage of the respondents and their preferences for specific features.

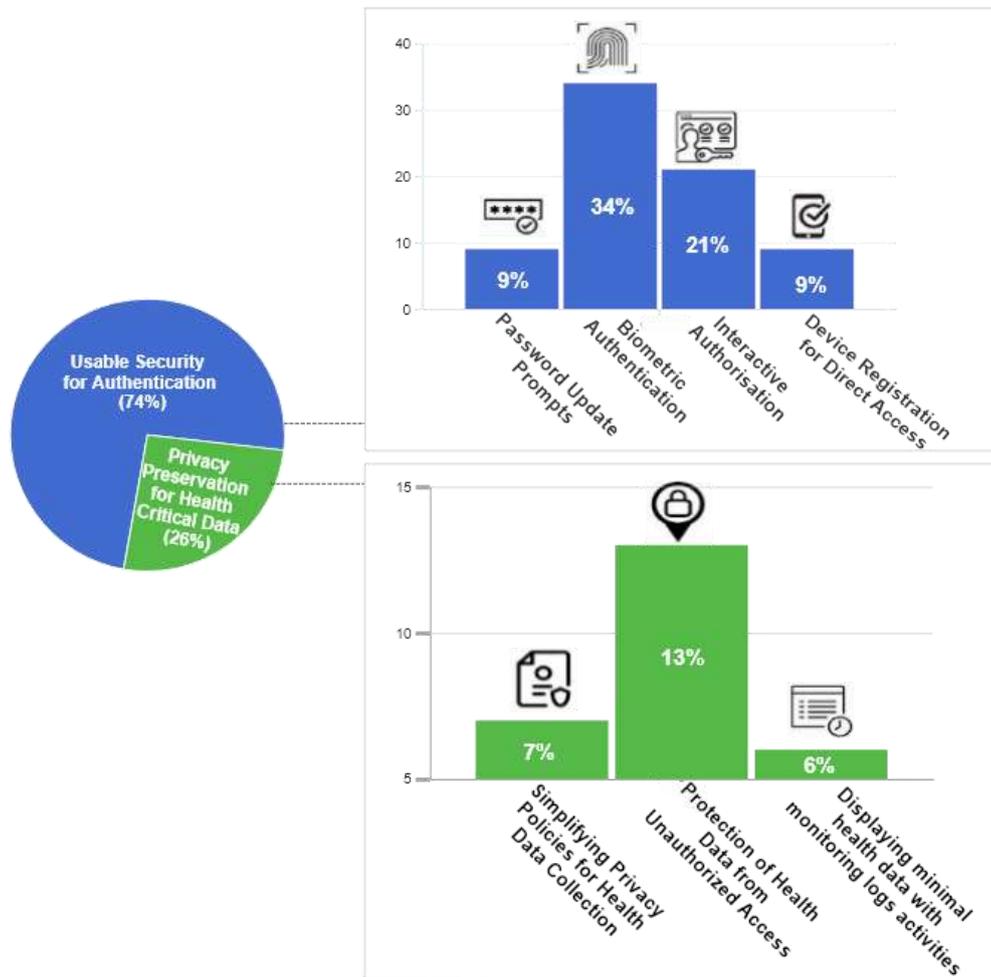

Figure 6. Preferred Security Features by our Respondents

*5.2.1. Usable Security for User Authentication*

As in Figure 6, 74% of the respondents desired a combination of secure and usable security features for authentication, referred to as usable security for authentication. Usable security refers to human aspects and their impacts on computer security, i.e., employing methods of human-computer interaction to support security features that are easy to use [51]. As a typical example, biometric verification is considered a usable security feature (i.e., enabling interactive, easy to use, personalized authorization) when compared with the traditional ID and code-based authorization. For detailed and fine-grained analysis of the desired features, we identified and represented five sub-features of usable security for authentication, each of which is detailed below and illustrated in Figure 6. Table 2 complements the discussion of **RQ-1** by highlighting some of the responses from end-users regarding their perception about existing and desired security features.

As in Figure 6, 34% of the respondents helped us to identify biometric verification as one of the desired features supporting usable security for authentication. For example, three of the respondents ([**R1**], [**R20**], [**R72**]) indicated their desired feature as: '*Accessing the app through facial recognition (becomes) easier and faster than using passwords*' [**R1**], '*Fingerprint to provide an additional feature for user verification*' [**R20**] and *'(I) prefer using a fingerprint to log in'* [**R72**]. We now present the identified sub-features of usable security for authentication, as illustrated in Figure 6.

− *Password Update Prompts:* It refers to a commonly implemented feature that frequently notifies the user to update their passwords and/or help them select passwords based on specific character combinations to increase password strength. As per Figure 6, 9% of the respondents suggested that mHealth apps should encourage end-users to change their passwords frequently and facilitate them about how to create strong passwords. For example, as in Table 2 the responses [**R28**] and [**R34**] indicated that the app should force users to regularly update their passwords to avoid password breaching.

− *Biometric Authentication*: It refers to exploiting the unique features of human biometrics such as fingerprints, facial imaging, or retina imprints to enable user authentication. Compared to others means of authorization and authentication**,** biometric authentication is the most desired feature suggested by 34% of the respondents. Some example responses such as [**R1**] and [**R19**] suggested that finger printing or facial recognition is an easier, faster, and foolproof way to log in to the mHealth app. *'Accessing the app through facial recognition is easier and faster than using passwords'* [**R1**], and *'Use of a fingerprint is helpful to login but sometimes the verification text arrives too late ...'* [**R19**].

– *Interactive Authorization:* It refers to prompting for further input to grant the user to access resources (e.g., 2FA) [52]. 21% of respondents preferred the employment of interactive authorization in mHealth apps. Respondents (e.g., [**R15**], [**R43**], [**R46**], [**R80**], [**R96**]) pointed out that mHealth apps should be accessible after verifying the user. *'The app should be accessible by sending a verification code to the registered phone number, similar to my banking app'* [**R43**].

– *Device Registration for Direct Access:* It refers to registering a device that can access the app installed on a device without any further authentication. This means that once the authorized user logs in to the device, he can directly access the mHealth app, a feature desired by 9% of the respondents. The users indicated that having a registered device makes it more convenient and faster to log in [**R8**], whereas [**R11**] suggested he/she finds repeated logins as inconvenient and exhaustive and prefers an authorized person to access any app directly once logged in to their device. *'Saving passwords and allow users to login directly because he or she would be the only one who uses the mobile'* [**R8**], and *'Is it not possible to make the app accessible without re-entering the password as long as the same device remains with the person?'* [**R11**].

> Based on user responses, we observed that the majority of end-users desired convenience and ease of use, and fewer steps to log into the examined apps and access their private health-critical data. Based on the percentage of users in Figure 6 and example data in Table 2, we can conclude that respondents on one hand desired simplifying the current authentication methods to access the app, and on the other hand, suggested further restrictions to access the app such as multi-step authentication. In fact, it can be a daunting task to satisfy all end-users requirements for data access. One possible solution, as indicated by a few respondents, is that an mHealth app should enable users to select and configure the authentication method they prefer. For example, the app should allow users to select whether they would like to log in such as, biometrics, or device registration as indicated by [**R28**, **R68**, **R83**, **R101**].

Table 2. Example of User Responses (Existing and Desired Features) in the Context of RQ-1

|  | **Example of End-Users' Responses** | **Response ID** |
|---|---|---|
| | A. **Existing Security Features** | |
| | **App Permissions** | |
| **SA-1** | 'It is unknown how health data is being handled. But we trust (the health provider) to keep the app secure' | [R68] |
| **SA-2** | 'the app is easy to use for authentication and trusted for not asking my personal information that is not relevant to my health issues and visit to health centre' | [R36] |
| **SA-3** | 'the app does not collect health data and not attached to other devices; all data came from hospital's database' | [R48] |
| | **User Authentication** | |
| **SA-4** | 'The two-authentication factor is enough to ensure security' | [R96] |
| **SA-5** | 'Provide some instructions for the users on how to create strong passwords' | [R60] |
| | **User Control** | |
| **SA-6** | 'the app allows me to prevent or permit accessing only my contact and camera' | [R65] |
| **SA-7** | '(I am) not aware of how (my) health data can be erased from a lost or stolen device' | [R11] |
| | **Feedback and Reporting** | |
| **SA-8** | 'The app contains an icon to report any issues related eservice including the app' | [R31] |
| | B. **Desired Security Features** | |
| | **Usable Security** | |
| **Password Update Prompts** | 'Forcing the user to change the password periodically so it becomes prone to (password) sniffing attacks and only the actual user would access the app' | [R28] |
| | '(the app) must support reminders to change the password frequently and accepting strong passwords only' | [R34] |
| **Biometric Authentication** | 'The use of the fingerprint is easier and convenient for me but sometimes the verification message arrives too late' | [R19] |
| | 'Accessing the app through facial recognition easier and faster than using passwords' | [R1] |
| **Interactive Authorization** | 'The app should be accessible by sending a verification code to the registered phone number. Similar to the banking apps' | [R43] |
| | 'Double confirmation of access password and text code via phone SMS' | [R46] |

| **Device Registration for Direct Access** | 'Saving passwords on a registered device and allow users to login directly (with a touch or click) because I would be the only one who uses the mobile after I login to my device' | [R8] |
|---|---|---|
| | 'Is it not possible to make the app accessible without re-entering the password as long as the same device remains with the person?' | [R11] |
| **Privacy Preservation** | | |
| **Simplifying Privacy Policies for Health Data Collection** | 'Privacy policy in the app needs to be simple and easy to read for users' | [R24] |
| | 'Al Habib App stores many personal data about the users. The privacy policy is not clear on how to deal with data' | [R67] |
| **Protection of Health Data from Unauthorized Access** | 'The user should be informed about any data access, Notify access alert through application or SMS' | [R64] |
| | 'The app should have complete confidentiality of the patient's health and other private data' | [R70] |
| **Displaying minimal health data with monitoring logs activities** | 'The app should not display too much health information when it is not useful to display' | [R52] |
| | 'They should introduce a facility to review my profile access, including the lab results with access time and the requester information' | [R64] |

*5.2.2. Privacy Preservation for Health-Critical Data*

Protecting the privacy of health-critical data and private information of users by employing various privacy preserving methods is desired by 26% of the users, as in Figure 6. For example, **R70** indicated, *'The app should have complete confidentiality of the patient's health and other private data'*. It should be noted that privacy preserving for health-critical data overlaps with the confidentiality in regards to unauthorized access, health data disclosure, sharing health data [25]. A few respondents were concerned about who accesses their health data (i.e., authorization). *'Notify access alert through application or SMS'* [**R64**].

Our analysis of the survey responses revealed three features to preserve user privacy in mHealth apps, each detailed below.

– *Simplifying Privacy Policies for Health Data Collection:* Privacy policies refer to statements or a legal formality that details how a mobile app will collect, process, and exchange users' personal information. Privacy policies help users to understand the potential for manipulation of their classified information by an app and any control provided to allow or restrict such access. As per Figure 6, 7% of the respondents (e.g., [**R24**], [**R67**]) expressed their concern about the privacy of health data as the privacy policy is not clear in their opinion and hence, they suggested simplifying privacy policies so they could be easily understood. For example, the response [**R67**] suggested that the privacy policy is complex and that it is hard to understand how mHealth apps handle health-critical data. *'(the app) stores many types of personal data about the users. Privacy policy is not clear on how to deal with such data'* [**R67**].

– *Protection of Health Data from Unauthorized Access:* Privacy of health data ensures that classified data is kept private and only shared with entities (e.g., human or computer programs etc.) that are authorized to access it. 13% of the respondents, such as [**R64**], and [**R70**] indicated the features to ensure the privacy of their health data and private information. [**R64**] suggested the use of proper notification or alerts in the case of private information being accessed by any third party. *'Notify access alert through application or SMS'* [**R64**].

- *Displaying minimal health data with monitoring logs activities:* Minimal data display ensures that only the most relevant information is shown via an app to even to the authorized person viewing it. This means that a clinical technician that needs to pass on a patient's information to a doctor must not be provided with insights into a patient's health conditions and disease symptoms, unless explicitly required. In addition, monitoring logs (audit logs) helps to trace unusual access patterns that can be restricted in the future. Some of the respondents suggested a complementary audit logging feature to capture and review account activities. 6% of our respondents suggested displaying less health data and monitoring who accesses their data. For example, respondent [**R52**] suggested that mHealth app should not display too much health information when it is not required. *'The app should not display too much health information when it is not useful to display'* [**R52**]. Respondent [**R64**] indicated that the app should provide the facility to review a user's profile access time and the requester information. *'They should introduce a facility to review my profile access including the lab results with access time and the requester information'* [**R64**].

🗒️ We conclude that end-users were concerned about their privacy and suggested further mechanisms to preserve the privacy of their health-critical data, as shown in Figure 6 and example data in Table 2. The respondents highlighted some issues that needed to be considered to guarantee data privacy. The implementation of a few features was desired, including simplifying privacy policies for health data collection, and ensuring that health data is shared with authorized entities. Furthermore, minimal health data should be displayed and log activities monitored to trace abnormal data access.

## 6. Security Related Issues in mHealth Apps (RQ2)

We now answer **RQ2** which aims to investigate the security issues faced by end-users during their usage of the employed security features in the investigated mHealth apps (i.e., **Q11**, per survey design in Figure 2, Phase 1). In our context, we define a security issue as a problem or a challenge experienced by end-users related to security features or other aspects of the app while using the app; for example, requesting permission to access more device resources (e.g., mic or camera) or data (e.g., contacts or photos) than the app actually needs. Based on the collected responses, we identified a multitude of security-related issues ranging from concerns about multi-step authentication to discomfort with sharing health-critical information with stakeholders. The reported issues can provide guidelines to further optimize security and usability of mHealth apps. We also observed that several end-users were not able to point out any security-related issues due to their lack of knowledge (security perception **RQ1**). For example, one of the respondents indicated that: 📧 '*I have not encountered any (security) problems but my mHealth app provides appropriate security measures to protect my data*' [**R33**].

Some respondents indicated miscellaneous issues that can be classified as generic or performance-related issues such as 'app freezing' 'poor quality of medical imaging', 'lack of notifications' reported by the respondents [**R5**], [**R6**], [**R37**], [**R71**], [**R95**] which were not relevant to app security and were discarded during analysis (as in Figure 4). We classified the reported issues into three main categories as in Figure 7, each of which is detailed below.

### 6.1 Delays in Two-Factor Authentication (2FA)

Difficulty or delays in authenticating an app's users can be related to the issues in the execution of security procedures that are responsible for gathering users' credentials to verify and authenticate for system login. The primary issue reported in this context is 2FA that first collects users' credentials and then authenticates them via an SMS to the registered device or telephonic number.

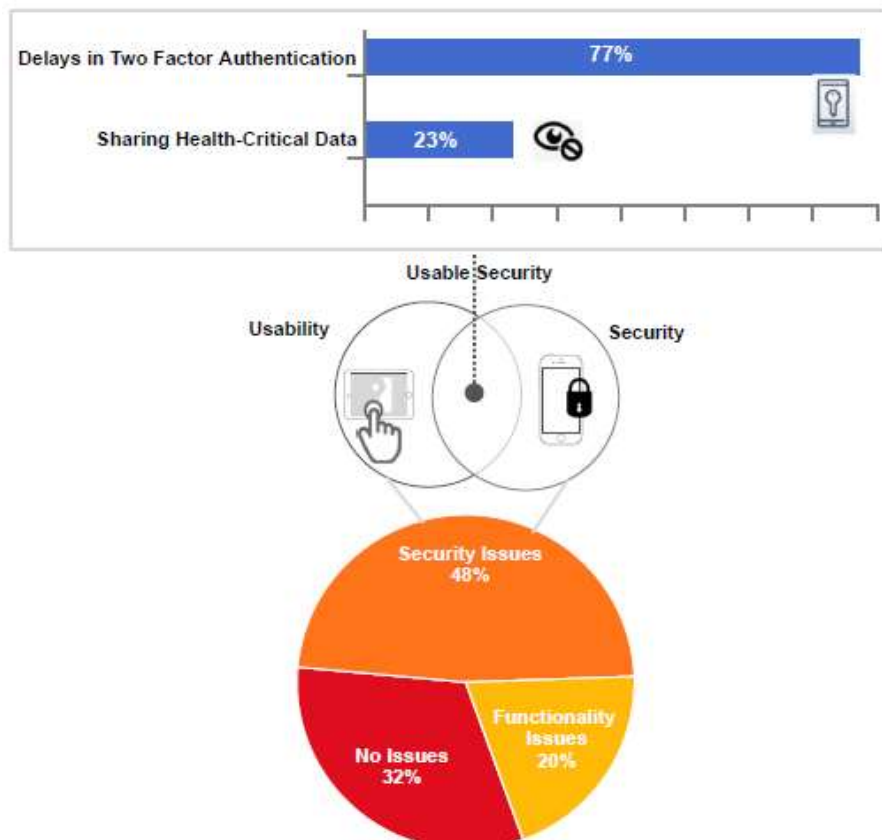

Figure 7. Security Issues with the Examined mHealth Apps reported by our Respondents

Security regulations and policies such as HIPAA and EU GDPR recommend the implementation of 2FA features for ensuring the security of mHealth apps by adding an extra layer of authentication on top of traditional identity and passcode-based access [27, 29]. However, the respondents indicated that such a delay in getting the verification code would affect the usability of the apps. 2FA enhances an app's security but can restrict a user from entering the app if the SMS is not delivered due to network or connectivity issues. For example, several respondents (e.g., [**R2**], [**R18**], [**R19**], [**R54**], [**R56**], [**R57**], [**R92**]) reported that the delays in getting the app verification code in some instances lead to session expiry. Specifically, two of the respondents suggested that, 🖥 '*Sometimes verification message is not received in a timely fashion and I cannot get (into) the app*" [**R18**], and, *"Due to weak network signals, sometimes I do not timely receive the SMS to change my password'* [**R43**].

### 6.2 Sharing Health-Critical Data with Stakeholders

Securing private data and ensuring its privacy from other humans or non-humans such as third-party programs that sense users' location and context is a critical concern for end-users. The survey findings indicate that some of the users are uncomfortable with the fact that all medics (manager level users of the app, e.g., nurses, doctors, technicians), regardless of their medical relevance, can search and view health-critical information of any patient. Furthermore, the access privileges for managerial level users of apps can be extended beyond the working hours and health unit premises. This raised concerns about privacy of users' health-critical data (disease symptoms, medical images etc.) that can be leaked and have specific social consequences. For example, respondent **R52** indicated that: 🖥 '*Health information should remain between doctor and patient and should not be shared outside the hospital. They (the medics) have the access to patients' data, and they can share my information with others without my consent or any alerts from the app*'. The user should be able to give consent to access health data as **R24** suggested. **R24** indicated: 🖥 '*The medical record can be accessed after obtaining approval from the user and it is in the form of a code sent as a text message on his/her mobile'*. Specific concerns for data privacy included using personal information for advertising purposes. [**R83**] indicated that: 🖥 '*The app should not display too much health information when it is not useful to display'*.

> 📝 Developing secure mHealth apps requires following a certain guideline, such as HIPAA, to ensure the security expectations are met. However, as illustrated in Figure 7 and example data in Table 6 (Appendix 2), some of our respondents found some of the employed features challenging. Our respondents indicated that 2FA through messaging a code via SMS can cause difficulties in accessing mHealth apps. Even though 2FA provides an extra layer of protection, it needs to be more convenient for end-users. Also, the respondents pointed out that health professionals should keep end-user's data confidential. It would be possible to obtain end-users' consents through mHealth apps once sharing health-critical data with other health professionals is required.

## 7. Methods to Improve Security Awareness (RQ3)

We now answer **RQ3**, which aims to identify the methods and practices adopted by end-users or supported by mHealth providers to improve the security awareness of end-users (i.e., **Q12,** per survey design in Figure 2, Phase 1). Earlier studies such as [4, 15] indicate that developing secure apps or adopting state-of-the-art security practices may not be sufficient to protect the classified data of end-users. End-users' awareness in terms of their knowledge, attitude, and ability to identify threats and adopt security-aware practices for their private and health-critical data is of prime importance [36, 53]. For example, despite advanced data encryption techniques for medical records (i.e., technical perspective), selection of a weak password scheme or infrequent updates of it (i.e., human perspective) may expose users' data to vulnerabilities [21, 23]. Therefore, educating users or employing methods that enhance their understanding of security is important for mHealth providers to avoid any socio-legal challenges of security breaches [22, 43]. Based on the survey responses, we have identified and classified the reported methods that help end-users to improve their security awareness. First, we have analyzed the text of the survey responses to identify two categories: (i) end-users that self-educate and (ii) end-users that need support from app providers to improve their security awareness. Figure 8 visualizes both of the categories for fine-grained discussion of methods that improve security-awareness of end-users. We excluded comments that were based purely on user awareness about app usability with no link to security.

### 7.1. Self-Education
As in Figure 8, 32% of end-users indicated that they self-educate to improve their security awareness about the mHealth app. For example, the respondents (e.g., [**R11**], [**R14**], [**R23**], [**R31**], [**R48**]) indicated that they relied on themselves to explore security features or read the apps' manuals. Based on the demography analysis (Figure 3), we observed that almost all of the respondents opting for self-education have at least a bachelor's degree with moderate to advanced knowledge of IT. For instance, 💬 *'I learned about the app by myself and by practicing'* [**R14**]. Moreover, some of the respondents (e.g., [**R23**], [**R31**], [**R44**], [**R50**], [**R56**]) indicated that they had not been guided about secure usage of the apps or about utilizing the specific security features of the app but they were motivated to learn about usability, functionality, and security. For example, the respondents [**R44**] and [**R68**] suggested that exploration of the app's functionality and excitement about learning features to secure their personal data helped them to learn about the security features provided by the app.

## 7.2. Support from App Providers

As in Figure 8, 68% of the end-users indicated the need for support from app providers to improve the security awareness of end-users. For example, some respondents including [**R21**], [**R60**] suggested further support from the app provider to educate users about protecting their data. For example, 💬 *'(the app provider should provide) some instructions to the users such as how to create strong passwords or manage app access permissions'* [**R60**].

Based on end-user responses, the methods provided by app providers can be generally divided into three main categories, each of which is detailed as below and illustrated in Figure 8.

- *Security Awareness via Social Media:* As in Figure 8, a total of 14% of end-users suggested that they become security-aware via social media. As per some respondents [**R12**], [**R16**], social media campaigns, documentary tutorials, and videos help end-users to view, share and engage in discussions regarding various aspects of the apps that they use. The respondents indicated that social media presence of the mHealth app providers helped them to get useful information and query needed clarifications. For example, three respondents (i.e., [**R12**], [**R16**], [**R71**]) indicated that posting on a prominent social media platform like Twitter and uploading a tutorial video on YouTube have helped them to learn about app security features. For example, 💬 *'I watched a short video online about how to securely setup (and configure) my app'* [**R12**].

- *Content Support from mHealth Provider:* Figure 8 suggests that an overwhelming majority of users (i.e., 71%) indicated that they rely on the content from the mHealth providers to improve their security awareness. mHealth providers provide different types of content such as guidance from app experts to be consulted, email, helpline and support. 22 respondents (e.g., [**R4**], [**R13**], [**R26**], [**R34**], [**R100**]) reported that they have been notified by the security features by health providers' channels, which can be classified into five main types

*a. Guidance from doctors or nurses (23% respondents)*: Some of the respondents (e.g., [**R4**], [**R6**], [**R15**], [**R4-6**], [**R9**]) indicated that the assigned doctors or nurses showed them how to use the app. 💬 *'My doctor advised me to download, and he showed me how to use the existing features'* [**R4**].

*b. Advice from staff (19% respondents):* The respondents (e.g., [**R26**], [**R29**]) in this category indicated that medical support staff such as reception or helpdesk personnel helped them to familiarize themselves with the usability and security features of the app. 💬 *'While booking an appointment, a brief introduction of the app was given to me including password creation and accessing my data securely'* [**R88**].

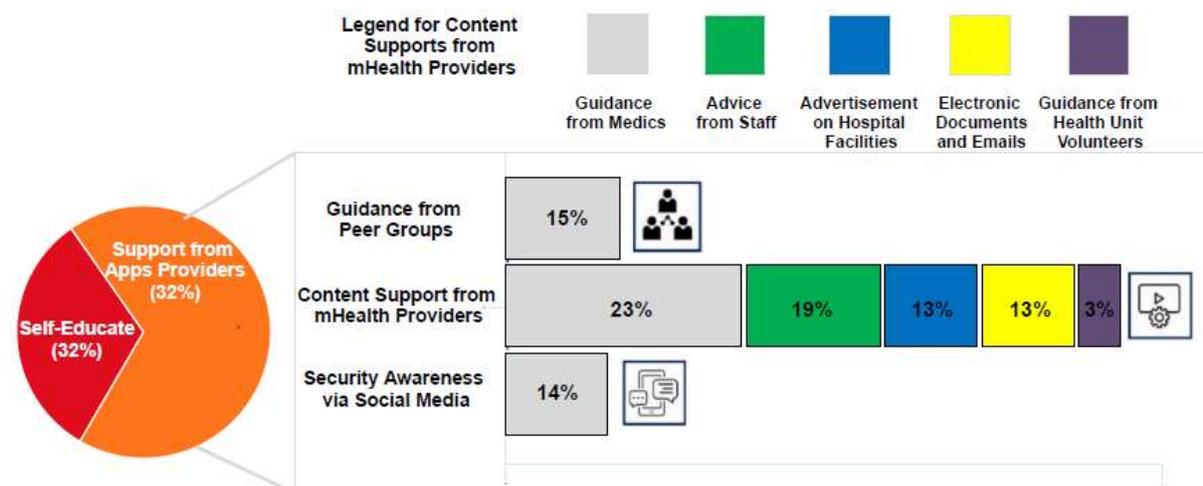

Figure 8. Methods to Improve End-users' Security Awareness by App Providers

*c. Advertisements on hospital facilities (13% respondents)*: The respondents (e.g., [**R17**], [**R55**]) indicated that the facilities and interaction on hospital premises helped understand the importance of securing health data and personal information. *'The interactive screens inside the buildings assisted me in learning about the app and made me realize about security of the data'* [**R17**].

*d. Electronic Documents and Emails (13% respondents):* The respondents including [**R10**], [**R34**] pointed out that they learned about the security features of the app based on materials shared via electronic mail. *'I learned the security features through the frequent e-mails which I receive from the health provider to guide me on how to use the app'* [**R10**].

*e. Guidance from Health Unit Volunteers (3% respondents):* Only a minority of the respondents (using KFMC app) indicated that meeting a volunteer helped them become aware of the security features. *'The app was presented to me […] through a volunteer team to help me understand its usage and basic security features'* [**R100**].

**-** *Guidance from Peer-Groups:* Peer groups of mHealth apps represent a group or circle of friends who use mHealth apps like family contacts or colleagues. Figure 8 highlights a total of 15% of end-users that rely on guidance from peer-groups for security awareness. Specifically, five respondents (e.g., [**R3**], [**R7**], [**R25**], [**R72**]) indicated that they have appropriate knowledge about using the app securely and it is observed that the improved skills are the result of a respondent's consultation with a friend, family member or colleague at work. *'My younger brother recommended and downloaded the app for me. He helped to sign up and showed me how to use it and keep it secure'* [**R25**].

> Based on the discussion above regarding the approaches to improve users' security-awareness, we conclude that end-users have not received sufficient awareness training on how to use mHealth apps in a secure way, as shown in Table 7 (Appendix 2) and Figure 8. Even though 83% of our respondents relied on the content from the mHealth providers, we argue that the objective was more about marketing the app rather than providing security awareness. We believe that end-users should be further assisted through providing security training. Such training should be delivered through security experts to ensure end-users become aware of security threats and the appropriate mechanisms to manage risks.

## 8. Discussion of the Key Findings

Our study empirically investigated the security awareness of end-users about using clinical mHealth apps. We presented our findings based on participants from our case health providers. While it can be argued that some of the recommendations and guidelines presented in this study are applicable to other mobile apps, the outcomes represent the participants' views. Deciding what security features should be employed in mHealth apps rests upon health providers. We now discuss the key results that highlight the core findings for **RQ1** – **RQ3** based on the methodological steps in Figure 2, and outline the scope for future work. The discussion is guided by Table 3, which provides an illustrative summary as a taxonomy of the key results. Based on Table 3, first, we highlight the demography and app usability analysis (Section 8.1) followed by a summary of answers to RQs (Section 8.2 – Section 8.4). For example, Table 3 highlights that, in the context of **RQ1**, biometric authentication is a desired feature of usable security. As explained earlier, several respondents desired biometric authentication (e.g., facial recognition) as a more user-friendly and faster mechanism to access the app compared with typing their ID and password.

*8.1. Demography and App Usability Analysis*

The demography analysis helped us to investigate if factors like users' experience, education or IT proficiency impact their security awareness and reflect any patterns for secure usage. For example, the respondents such as [**R11**] and [**R14**] (who have bachelors' degrees and knowledge of IT) pointed out that their concerns about the security of their data motivated them to self-educate about the app's security (RQ-3). In comparison, some respondents such as [**R66**] (with a high school degree and minimal to no knowledge of IT skills) were not sure about the features offered by the app for data security. The study acknowledges that some of the received responses focused solely on app usability and security-specific issues, since most of the respondents lacked IT knowledge or experience with mobile app usage. Such analysis provides a fine-grained investigation of end-users to analyze if education, IT proficiency and usage history impacts end-users' awareness, and ultimately increases the security of their health-critical data.

Table 3. Taxonomical Classification of the Core Findings (Key Results for all RQs)

| App Usability Analysis | | | | | | |
|---|---|---|---|---|---|---|
| Mobile Platforms | Used app | Gender Classification | Age Group | IT Knowledge Level | Education Level | App Usage |
| • 33% Android<br>• 66% iOS<br>• 1% Meizu | • iKFMC App (62%)<br>• Dr. Sulaiman Al Habib app (38%) | • 41% Female<br>• 59% Male | • 33% 18 – 29 years<br>• 59% 30 – 49 years<br>• 8% Above 50 years | • 31% Little/no knowledge<br>• 48% Moderate knowledge<br>• 21% Advanced knowledge | • 11% High school or less<br>• 13% Diploma<br>• 45% Bachelor<br>• 10% Higher diploma<br>• 22% Master's or PhD | • 6% At least once a day<br>• 14% At least once a week<br>• 37% At least once a month<br>• 23% At least once every 3 months<br>• 14% At least once every 6 months<br>• 7% At least once a year |

| Security Perception (RQ1) | | | | | | |
|---|---|---|---|---|---|---|
| **Securing Health Data** | **Security Awareness** | | | | **Desired Features** | |
| The importance of securing health data within mHealth apps | App Permissions | User Authentication | User Control | Feedback and Reporting | Usable Security for Authentication (74%) | Privacy Preservation (26%) |
| • 87.1% believed it is very important<br>• 7.9% believed it is important<br>• 4% of respondents were neutral<br>• 1% believed it is not important | *User Consent:* **20.8%** were unaware if the apps requires users' consent or not.<br>*Information Access:* **40.6%** were unaware if the apps collect data without permission or not.<br>*Data Collection:* **61.4%** knew that the apps don't collect more personal information. | *2-Step Authentication:* **24.8%** were unaware if the apps support 2-step authentication or not.<br><br>*Password Strength:* **35.6%** were unaware if the apps accept a weak password or not. | *Adjustable Security:* **48.8%** were unaware if the apps have adjustable security or not.<br><br>*Data Wiping:* **60.4%** knew that the apps have the data wiping feature. | *Backend Security:* **46.5%** were unaware if the apps have the feature of reporting security issues or not. | • Password Updates (9%)<br>• Biometric Authentication (34%)<br>• Interactive Authorization (21%)<br>• Device Registration for Direct Access (9%). | • Simplifying Privacy Policies (7%)<br>• Protection of Health Data from Unauthorized Access (13%)<br>• Displaying minimal health data with monitoring logs activities (6%) |

| Security Issues (RQ2) | Security Education (RQ3) |
|---|---|
| **Security issues that end-users faced during their usage of the security features within mHealth apps.**<br>• Delays in authentication (77%)<br>• Sharing their health data (23%) | **Methods that improve the security knowledge of end-users towards using mHealth apps.**<br>• Self-education (32%)<br>• Support from app providers (68%) as follows:<br>  - Guidance from peer groups (15%)<br>  - Security awareness via social media (14%)<br>  - Content support from mHealth provider (71%) including<br>    a. Guidance from doctors or nurses (23%)<br>    b. Advice from staff (19%)<br>    c. Advertisements on hospital facilities (13%)<br>    d. Electronic Documents and Emails (13%)<br>    e. Guidance from Health Unit Volunteers (3%) |

Based on the survey responses, 95% of our respondents believed that the security of their private information (e.g., location, age, gender) and health data (e.g., blood pressure, medical history) is a critical concern to them. Specifically, 88 respondents (i.e., 87.1%) selected very important, 8 of them selected important (7.9%), and a further 4 (4%) remained neutral, as highlighted in Figure 9. Surprisingly, one of the respondents indicated that security of mHealth apps is not important at all as (s)he considers it valuable if the app enables her/him to share his/her health and fitness routine with his social media contacts. Specifically, the respondent [**R8**] suggested that

*'I use (...) app for consulting (the nutritionist) regarding my diet and fitness monitoring and security of my (dietary, exercise, fitness monitoring, etc.) data is not very critical. I would prefer if it allows me to share my workout, diet plans and recommendations from the doctor (i.e., nutritionist) can be read and automatically shared with my (social media) friends and contacts. I will not be comfortable sharing my (exact) location but other than that if the app shares my (fitness) data with my permission, I am OK with that.'*

> The findings indicated that the vast majority of our respondents (95%) are concerned about securing their health-critical data. Only 1% was less concerned about health-critical data that corresponds to health and fitness monitoring.

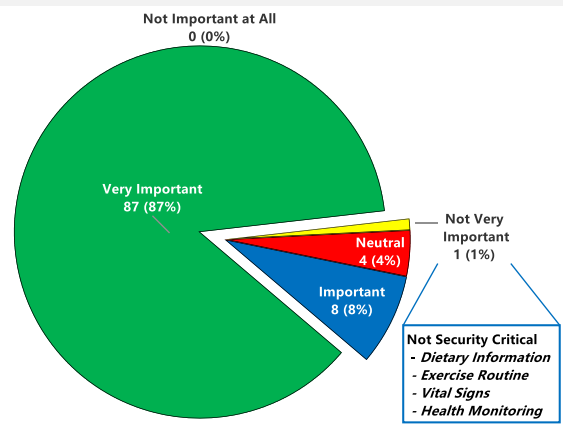

Figure 9. Respondents' Perceived Importance of Securing Private Data within the Apps

### 8.2. End-users' Understanding of Existing Security Features and the Desired Security Features

Security perception of end-users is indicated based on their understanding of the existing features (Figure 5) and the knowledge about the desired features (Figure 6) that enable or enhance app security. The key findings indicate that the majority of the end-users were unaware of the security features that had been implemented in the app they were using (Figure 5). For example, 26.7% of our respondents were aware that the mHealth apps have very adjustable security settings that help to control apps permissions. The findings suggest that end-users find it impractical to have apps that offer a multitude of security features that are difficult to understand or use.

The respondents indicated a total of seven desirable features that support (i) usable security for authentication (i.e., 74% end-users) and (ii) privacy preservation (i.e., 26% end-users), as illustrated in Figure 6. For example, Device Registration for direct access to app data is a desired feature as indicated by [**R43**]. 43% of the end-users liked the security features such as biometric authentication, or device registration for user authentication.

> Our study revealed that the respondents did have concerns about the security and privacy of their health-critical data. At the same time, we found out that the majority of our respondents are still not fully aware of the employed security features that help to control mHealth apps. In fact, **R21** suggested that there is a need to raise awareness about using mHealth apps. Because mobile devices are more vulnerable to security breaches, there is a clear need to ensure that end-users are familiar with basic security features which protect their critical data in order to enhance app security, and increase and increase users' trust in using the app. In particular, the use of BYOD has the potential to damage/ compromise security since mobile devices run other mobile apps, which frequently access device data, in parallel to mHealth apps. As a result, lack of security awareness can compromise the security and privacy of health-critical data. We also identified seven security and privacy-related features based on the surveyed end-users. Our findings can be useful for mHealth app developers to consider the identified security features as guidelines and recommendations. More importantly, convenient apps need to ensure a tradeoff between security and privacy, on the one hand, and mHealth apps usability on the other hand.

*8.3. Security Specific Issues Faced by End-Users*

The respondents were asked about the security-related issues that they experienced while using the security features in the examined mHealth apps. Our study confirmed that the examined mHealth apps still have some security issues that our respondents have experienced. Our findings are consistent with those of previous studies (e.g., [22, 23, 38]) regarding the security issues faced by end-users. As in Figure 7, 48% of the respondents reported usable security issues faced by them that affect user experience or discomfort about app usage. Implementing the 2FA to access mHealth apps aligns with one of the recommended security practices by health regulations (e.g., HIPAA). However, it limits end-users to a specific authentication method, which sometimes does not work well (e.g., due to weak cellular network signals) and can affect the availability of the apps. Another security concern mentioned by our respondents is violation of end-user's privacy (Figure 7) by managerial level users of the app e.g., nurses, doctors, technicians. Some respondents indicated that medics can access and share their health-critical data without getting consent from end-users.

> In summary, our findings presented end-users' evaluation and suggest a few security measures that would provide more secure, effective, and convenient mHealth apps. First, issues during authentication were the most indicated by our respondents, and better authentication methods, the most desired features. Therefore, we conclude that following HIPAA regulations, i.e., selecting two of the three recommended authentication methods: i) something an end-user knows (e.g., ID and passwords), ii) something an end-user owns (e.g., authentication tokens), or iii) something an end-user finds most convenient (e.g., biometric authentication) would increase end-users' trust in using mHealth apps, and enhance the security. On the other hand, ensuring the privacy and confidentiality of users' health data was a big concern for a few respondents. For instance, unnecessary authorization, especially for different units of the health providers, can lead to exposure to health data. Thus, suitable privacy-preserving methods, such as anonymization, could be employed by the developers to enhance end-users' use of mHealth apps. Furthermore, health data should be subject to a proper access control policy (i.e., notifying end-users who accesses their health-critical data, when, and for what reasons). We believe our results can help guide the development team to incorporate better security features.

*8.4. Methods to Improve Security Awareness of End-Users*

Raising the security awareness level of end-users towards using mHealth apps is crucial for health providers to avoid security risks. As illustrated in Figure 2, **RQ3** aimed at investigating how end-users of mHealth apps become aware of the security features offered by the apps. On the one hand, 32% of our respondents (e.g., [**R11**], [**R14**], [**R23**]) preferred self-education to improve their security perception (Figure 8). Demography analysis suggested that almost all such respondents that advocate self-learning and education have an appropriate level of IT knowledge or they have installed and used similar apps (e.g., other health provider apps) on their devices. On the other hand, 68% required help and support via social media (e.g., [**R12**], [**R16**]), through the content provided by the health providers (e.g., [**R4**], [**R26**]), and/or needed guidance from peer-groups (e.g., [**R3**], [**R25**]) to improve their security awareness. Security awareness towards using mHealth apps is getting the attention of end-users and they welcome any opportunity to be educated.

> To summarize, providing security awareness for end-users is just as important as developing secure mHealth apps. There is a need for suitable security awareness methods supported by internal security experts to help end-users to avoid security-related risks. 42% of the respondents, such as [**R9**] [**R15**] [**R26**] indicated that medics or staff members had helped them in some way to become more aware of the apps, as in Figure 8. It would be quite helpful to organize short sessions to enable medics and staff members to be fully aware of all security aspects of the provided apps to ensure that end users are educated with appropriate security awareness. Such training could be arranged through a guidance program with each unit assigning a person to teach others within the unit to overcome the large numbers of participants.

## 9. Threat to Validity

We now discuss some of the threats to the validity of this study. There are three types (i.e., internal, construct, and external) validity threats to be discussed below.

*9.1. Internal Validity*

Internal validity refers to the extent to which the observed results were from a reliable population. The proposed study collected data by means of surveying end-users of mHealth apps. Relying solely on the end-user survey for data collection and lack of triangulation could have impacted the reliability of the findings. Another possible threat relates to the participants' subjective viewpoints, which could have been misinterpreted when analyzing the qualitative data. To overcome this threat, we performed (i) initial coding of the data (by the first author) followed by (ii) evaluation and finalization of the codes (by the second author). Figure 4 in Section 4.3 presents an example of data coding as an attempt to minimize this threat.

*9.2. Construct Validity*

Construct validity refers to the extent to which the used instrument measures what it is supposed to be measured. Employing the suitable instrument for data collection and synthesis can threaten the validity of study results. To ensure un-biased data collection, we designed an initial version and conducted a pilot survey (engaging 10 participants, Figure 2) based on the literature review and then analyzed the mHealth apps that were deployed on-site by our collaborators (KFMC and HMG). The pilot survey helped us to revise the survey questionnaire by eliminating any bias or confusion in the survey statements. The respondents who participated in the pilot survey suggested clarifying adjustable security settings in the sixth statement of **Q9**. Thus, we updated the relevant part of the survey with the addition of a controlling app permissions option, as in Figure 2. Furthermore, the initial questionnaire and pilot survey were validated by other members within the research group. Due to data privacy and ethics approval, the survey was possible only within the premises of our collaborating mHealth providers. As a result, it was not possible to collect data from other mHealth providers. This limited our efforts to engage more end-users with diverse backgrounds and different app usage, beyond two healthcare providers. The end-user survey could not accommodate the other means of data collection such as users' interviews and focus groups interactions for fine-grained collection and analysis of security awareness.

*9.3. External Validity*

External validity refers to applying generalization of the study results. As per the statistics for the number of downloads, according to Google Play store (App Store does not show publicly the number of downloads it can be viewed by app providers only), each app (i.e., KFMC app, HMG app) was downloaded by more than 10,000 end-users. In comparison to on-site data collection, a web-based survey with geographically distributed end-users can increase the participation (number of participants) and diversity of data collection (different apps from across the globe). The findings of our study are based on end-users' views from two health providers in Saudi Arabia; and hence, our results may not be generalizable due to geographical restrictions. Furthermore, our respondents might be influenced by the assigned policies and regulations, when following /and follow practices in the examined mHealth providers in Saudi Arabia. Therefore, we plan to further extend this research by collaborating with other mHealth providers globally.

## 10. Conclusions and Future Work

mHealth apps have been gaining more attention recently because they provide innovative solutions to deliver health services. However, despite their promising benefits, mHealth apps are susceptible to many security threats that jeopardize end-users' health-critical data and personal information. We conducted end-users' survey-driven case study research to understand the security knowledge and perceptions of the end-users of mHealth apps. We used a survey method to collect, analyze, and document the end-users' responses. The survey questionnaire was completed by 101 respondents, who were using mHealth apps provided by two approved mHealth providers in Saudi Arabia. Our data collection was enhanced by the demography data of the respondents, which helped us to highlight the supporting factors such as level of IT knowledge, age group, past experience with mHealth apps, mobile platforms they use, and educational backgrounds that increase our respondent's security-awareness. The key findings of our study are:
- Respondents had significant variances in their knowledge about the existence of the security-related features in the investigated apps.
- Most of the desired security features are related to usable security for authentication and preserving privacy.
- Difficulty in authenticating users and sharing health data were the reported security issues by our respondents.
- Security awareness towards secure usage of mHealth apps by security experts is missing. Yet, some respondents reported that health providers have supported them in some way/ to some extent to understand the apps.

This research investigated human-centric knowledge based on empirical evidence and provides a set of guidelines to develop secure and usable mHealth apps. We believe our study uncovered a few implications for future work including:

(i) Helping end-users to understand and prevent any security-related risks while using the apps. Furthermore, the developed security policy and guidelines could clarify the right actions that end-users need to take in different circumstances. At the same time, providing suitable security awareness regarding the policy and guidelines for end-users is as important as developing secure mHealth apps.

(ii) Our study can be further extended to investigate the impact of the desired security features and how that would lead to more secure mHealth apps.

(iii) Future research could investigate the impact of employing strict security features (e.g., two factor authentication). Such an investigation would identify impractical security measures, that can be further improved to present usable and secure mHealth apps.

The results of this study can benefit:
- Researchers who are interested in exploring human-centric knowledge for secure development and usage of mobile health applications. The key results streamline potential areas of futuristic research and a new hypothesis to be tested in the context of security vs usability of mHealth apps.
- mHealth apps developers and/or stakeholders interested in the fine-grained analysis of security features desired by end-users. In particular, the end-users' perspective could help developers to engineer next generation of mHealth apps that are secure and usable.

**ACKNOWLEDGMENTS**

We thank the research centers at KFMC and HMG for approving our study. We also want to thank all respondents for being a part of our study. Lastly, we thank the anonymous reviewers for their valuable comments/suggestions that improve the paper. This work is partially funded by Cyber Security Cooperative Research Center.

**Appendix 1. Respondents' Security Awareness Based on Demography Information**

As indicated in the Research Design (Phase 3 – Perform Data Analysis), we conducted *Independent-Sample T test* for gender since we were comparing two independent populations (i.e., male and female), and *Kruskall-Wallis H test* for more than two independent populations (e.g., IT knowledge level, age group, etc.). These statistical tests helped us to determine whether there were statistically significant differences on the level of security awareness among the defined groups of users. We further investigated the significant differences for the groups whenever applicable (i.e., whenever p-value < .005). We used the Mann Whitney U test to compare the median differences between the overall extents [54, 55]. For each demographic data, we tested the null hypothesis (i.e., $H_0$: *there is no significant difference*) against the alternative hypothesis (i.e., $H_1$: *there is a significant difference*), whereas $\mu_1$, $\mu_2, ..., \mu_k$ refers to population means.

Table 4. Variables and corresponding codes in SPSS

| Security Awareness | Code | Gender | Code | IT Knowledge Level | Code | Age Group | Code | Formal Education | Code | Frequency of Usage | Code |
|---|---|---|---|---|---|---|---|---|---|---|---|
| *Never* | 1 | *Male* | 1 | *Little or no knowledge* | 1 | *18 – 29 young adult* | 1 | *High school or less* | 1 | *At least once a day* | 1 |
| *Rarely* | 2 | | | | | | | | | *At least once a week* | 2 |
| *I don't know* | 3 | *Female* | 2 | *Moderate knowledge* | 2 | *30 – 49 adult* | 2 | *Diploma* | 2 | *At least once a month* | 3 |
| | | | | | | | | *Bachelor degree* | 3 | *At least once every 3 months* | 4 |
| *Sometimes* | 4 | | | *Advanced knowledge* | 3 | *Above 50 senior* | 3 | *Higher diploma* | 4 | *At least once every 6 months* | 5 |
| *Always* | 5 | | | | | | | *Master's or PhD* | 5 | *At least once a year* | 6 |

Table 5. Differences in Security Awareness Based on the Characteristics of Study Respondents

| **Demography Data Category** | **Identified groups** | **N (%)** | **p-value** |
|---|---|---|---|
| *Gender* | Male | 61 (59%) | .961 |
| | Female | 40 (41%) | |
| IT Knowledge Level | Little or no knowledge | 31 (31%) | .006 |
| | Moderate knowledge | 49 (48%) | |
| | Advanced knowledge | 21 (21%) | |
| Age Group | 18 – 29 young adult | 33 (33%) | .141 |
| | 30 – 49 adult | 60 (59%) | |
| | Above 50 senior | 8 (8%) | |
| Formal Education | High school or less | 11 (11%) | |
| | *Diploma* | 13 (13%) | .007 |
| | *Bachelor degree* | 45 (45%) | |
| | *Higher diploma* | 10 (10%) | |
| | *Master's or PhD* | 22 (22%) | |
| Frequency of Usage | *At least once a day* | 6 (6%) | .797 |
| | *At least once a week* | 14 (14%) | |
| | *At least once a month* | 37 (37%) | |
| | *At least once every 3 months* | 23 (23%) | |
| | *At least once every 6 months* | 14 (14%) | |
| | *At least once a year* | 7 (7%) | |

*1. Security Awareness based on gender*

To understand and compare the security awareness about the existing security features, we performed a statistical test (i.e., Independent Sample T test) to show if there was a significant difference between male (n=61) and female (n=40) respondents. The result indicated that there was no significant difference (i.e., variances are equal, $H_0$: $\mu_1$= $\mu_2$) between male (M = 3.627, SD =.766) and female (M = 3.634, SD =.638) (t= -.049, p-value= .961). Thus, we concluded that both males and females in our sample have equal security awareness towards the existing features.

*2. Security Awareness based on IT knowledge level*

We conducted the Kruskall-Wallis H test to examine if there were any significant differences in the security awareness of the existing security features among the three groups (i.e., *Little or no knowledge*, *Moderate knowledge*, and *Advanced knowledge*) as in Table 5. Our findings suggest that security awareness differed significantly (p=0.006). For the *Advanced IT knowledge*, the mean rank = 33.81 which is less than the mean rank = 52.78 for *Moderate IT knowledge* and less than *Little or no knowledge* (mean rank = 59.84). Specifically, significant differences were found (using the post-hoc Mann-Whitney U Test) between the *Little or no knowledge* group compared with the *Advanced knowledge* group (p =.003). In addition, we found that there was a statistically significant difference in the security awareness between the *Moderate knowledge* group, and the *Advanced knowledge* group (p=.010). On the other hand, we found that there was no statistically significant difference in the security awareness between the *Little or no knowledge* group, and the *Moderate knowledge* group (p=.256). Overall, we observed that end-users with *Advanced knowledge* of IT had higher security awareness scores, compared to those who had *Moderate knowledge*, or *Little or no knowledge*. Therefore, we reject $H_0$ and accept $H_1$ concluding that IT knowledge level had an impact on the security awareness of our respondents.

*3. Security Awareness based on age group*

We conducted a Kruskall-Wallis H test to determine any significant difference among the three age groups (i.e., n=33; young adult, n=60; adult, n=8; senior) in terms of their security awareness. To justify the low number of the senior sample (8 out of 101), we noticed during data collection that the majority of seniors (i.e., those above 50) do not use the provided mHealth apps for some reasons (e.g., they did not carry smart phones, even if they have smart devices it is not mandatory to use mHealth apps). Our findings suggest that security awareness for the groups, *adult*, *young adult*, and *senior adult* were mean rank = 46.87, mean rank= 54.85, and mean rank = 66.13 respectively (p=0.14). Since the *young adult* group had less security awareness than the *adult group*, we investigated the IT knowledge level for the *young adult* group (n=33) to elaborate a little bit on these results. We found that 30, i.e., 91% of respondents considered their IT knowledge as either *moderate knowledge* or *little or no knowledge*. Therefore, we accept H0: $\mu_1 = \mu_2 = \mu_3$ and conclude that age has no impact on the security awareness of our respondents.

*4. Security Awareness based on level of formal education*

We also wanted to investigate respondents' differences by considering the impact of the level of formal education on our respondents. We conducted a Kruskall-Wallis H test on the five groups, which we identified as shown in Table 5. Our results indicated that there is evidence (p=0.007) that security awareness of those with a postgraduate qualification (*Higher Diploma*, *Master's or PhD*) was lower, in terms of the sum of rank orderings, than those with an education level of *Diploma/Bachelor* less than *High School or less*. To further understand the difference within the five levels of education, we conducted a post-hoc Mann-Whitney U test. We found significant statistical differences in the security awareness between the *High school or less* and *Higher Diploma* groups (p= .005) and between the *High school or less* and *Master's or PhD* groups (p= .007). Further, we noticed a statistically significant difference between the *Bachelor degree* group and *Higher Diploma* group (p= .011). Our analysis also indicated a statistically significant difference between the *Bachelor degree* group and *Master's or PhD* group (p= .024). Therefore, we reject $H_0: \mu_1 = \mu_2 = \mu_3 = \mu_4 = \mu_5$ and conclude that the level of formal education has an impact on the security awareness of our respondents.

*5. Security Awareness based on frequency of mHealth app usage*

Lastly, we wanted to examine if our respondents' security awareness differed based on their frequency of usage. We conducted a Kruskall-Wallis H test on the obtained responses which were divided into six groups, as in Table 5. Our findings indicate that there was no statistically significant difference in the security awareness based on the frequency of mHealth app usage (p=.797). Similarly, we examined the security awareness based on the frequency of mHealth app usage through a post-hoc Mann-Whitney U test. We did not observed any significant differences among the six groups. Therefore, we accept $H_0: \mu_1 = \mu_2 = \mu_3 = \mu_4 = \mu_5 = \mu_6$ and conclude that the usage frequency for mHealth apps has no impact on the respondents' security awareness.

**Appendix 2.**

**Example of User Responses for RQ-2, and RQ-3**

Table 6. Example of User Responses (Security Related Issues) in the Context of RQ-2

| **Example of End-Users' Responses** | | **Response ID** |
|---|---|---|
| A. Delays in Two Factor Authentication | | |
| *'Sometimes verification messages not received in a timely fashion and I cannot get (into) the app'* | | **[R18]** |
| *'Due to weak network signals, sometimes I do not timely receive the SMS to change my password'* | | **[R43]** |
| B. Sharing Health-Critical Data with Stakeholders | | |
| *'Health information should remain between doctor and patient and should not be shared outside the hospital. Doctors have access to their patients data and they can share my information with others'* | | **[R52]** |
| *'My health information should not be disclosed without my permission and should not be shared with private clinics for the purpose of sending offers for me'* | | **[R83]** |

Table 7. Example of User Responses (Methods to Improve Security Awareness) in the Context of RQ-3

| **Example of End-Users' Responses** | | **Response ID** |
|---|---|---|
| A. Self-Education to Improve Security | | |
| *'The app is easy-to-use and does not need much awareness about its use and steps for changing my password or deny (requested) permissions'* | | **[R11]** |
| B. Support from App Providers to Improve Security Awareness | | |
| **Security Awareness via Social Media** | | |
| *'I watched a short video on Youtube about how to securely setup (and configure) my app'* | | **[R12]** |
| **Contents Provided by Health Provider** | | |
| **Guidance from doctor or nurse** | *'My doctor advised me to download, and he showed me how to use the existing features'* | **[R4]** |
| **Advice from staff** | *'While booking an appointment, a brief introduction of the app was given to me including password creation and accessing my data securely'* | **[R88]** |
| **Advertisements on hospital facilities** | *'The interactive screens inside the buildings assisted me in learning about the app and made me realize about security of the data'* | **[R17]** |
| **Receiving Support E-mails** | *'I learned the security features through the frequent e-mails which I receive from the health provider to guide me on how to use the app'* | **[R10]** |
| **Support from hospital volunteer** | *'The app was presented to me [...] through a volunteer team to help me understand its usage and basic security features'* | **[R100]** |
| **Guidance from Peer-Groups** | | |
| *'My younger brother recommended and downloaded the app for me. He helped to sign up and showed me how to use it and keep it secure'* | | **[R25]** |

**Appendix 3.**

**Survey Questions**

*Part 1: Demographic questions of the participants*

Q1: What is your age?   ☐ 18 - 29       ☐ 30 - 49       ☐ Over 50

Q2: What is your gender?   ☐ Female       ☐ Male       ☐ Prefer not to say

Q3: What is your level of education? ☐ High school or less       ☐ Diploma       ☐ Bachelor's       ☐ Higher Diploma   ☐ Higher education (Master's or PhD)

Q4: What is mobile OS?   ☐ I use Android (e.g., Samsung)       ☐ I use iOS (i.e., Apple)       ☐ Others, please specify

Q5: Which apps you're using?   ☐ iKFMC app       ☐ HMG app

Q6: Please rate you knowledge in using information technology? ☐ Little or no knowledge at all    ☐ Moderate knowledge   ☐ Advanced knowledge

*Part 2: Security issues of using mobile health apps*

Q7: What is the importance of securing your health data within the app? Your level of agreement will be measured by the following options:  Very important, Important, Neutral, Not very important, Not important at all.

Q8: Please respond to the following statements based on your experience with the mobile health application. Your level of agreement will be measured by the following options:  Always, Sometimes, Rarely, Never, I don't know

| # | Statement |
|---|---|
| 1. | The app has very clear, readable and understandable privacy policy. |
| 2. | The app requested my consent to share my health data. |
| 3. | The app does not ask for more personal information than what is needed. |
| 4. | The apps provide convenient options for authentications that support my needs. |
| 5. | The app does not collect data without my permission. |
| 6. | The app has very adjustable security settings and easy-to-use. |
| 7. | The app provides a channel to contact the developer or admin to report an issue. |
| 8. | The app has the feature of wiping all my health data if my phone is lost or stolen. |

Q9: What other security features you would like to have in mHealth apps?

Q10: What are the security barriers which you have experienced while using mobile health app?

Q11: What methods have been used to make you aware of the security features of mHealth app?